\documentclass[10pt,twocolumn,aps,prd,preprintnumbers,showpacs,superscriptaddress,nofootinbib,amsmath,amssymb,floats,floatfix,showkeys,notitlepage,longbibliography]{revtex4-2}
\usepackage{orcidlink}
\usepackage{comment}
\usepackage{lipsum}
\usepackage{graphicx}
\usepackage{subfigure}
\usepackage{palatino}
\usepackage{sans}
\usepackage{mathrsfs}
\usepackage{hyperref}
\hypersetup{colorlinks=true,linkcolor=blue,urlcolor=blue,citecolor=blue}
\usepackage[toc,page]{appendix}
\usepackage[normalem]{ulem}
\usepackage{adjustbox}
\usepackage{latexsym}
\usepackage{amsmath}
\numberwithin{equation}{section}
\usepackage{amssymb}
\usepackage{amsfonts}
\usepackage{dcolumn}
\usepackage{bm}
\usepackage{tikz}
\usetikzlibrary{decorations.pathmorphing}
\usepackage{bigints}
\usepackage{array,tabularx,multirow,booktabs}
\usepackage[tracking=true]{microtype}
\usepackage{soul} 
\SetTracking{}{500}
\SetTracking{encoding={*}, shape=sc}{40}
\UseRawInputEncoding 
\allowdisplaybreaks
\usepackage[utf8]{inputenc}
\usepackage{xcolor} 

\newcommand{\nn}{\nonumber\\}

\begin{document} \sloppy

\title{An obstruction and residual-completion theory for Newman--Janis deformations}

\author{Reggie C. Pantig \orcidlink{0000-0002-3101-8591}} 
\email{rcpantig@mapua.edu.ph}
\affiliation{Physics Department, School of Foundational Studies and Education, Map\'ua University, 658 Muralla St., Intramuros, Manila 1002, Philippines.}

\author{Ali \"Ovg\"un \orcidlink{0000-0002-9889-342X}}
\email{ali.ovgun@emu.edu.tr}
\affiliation{Physics Department, Eastern Mediterranean University, Famagusta, 99628 North
Cyprus via Mersin 10, Turkiye.}

\begin{abstract}
We formulate an obstruction and residual-completion theory for Newman--Janis-type deformations of black-hole seeds, where \textit{Newman--Janis-type} includes both the original complex-coordinate Newman--Janis algorithm and modified prescriptions such as non-complexification variants. The Newman--Janis algorithm is treated as an off-shell map from a static seed to a stationary-axisymmetric trial geometry, rather than as a solution-generating theorem. Its failure is encoded in a Newman--Janis obstruction tensor, defined as the residual obtained after substituting the trial configuration into the intended field equations. We decompose this residual into dynamical, geometrical/coordinate-admissibility, and source-preservation channels, separating field-equation failure from circularity, Boyer--Lindquist integrability, stress-tensor type, equation-of-state preservation, and matter-model realizability. When the obstruction is nonzero, residual completion asks whether a minimal correction of the metric and matter fields can cancel it. At leading nonzero order in the rotation parameter, this becomes a linear solvability problem: the obstruction must lie in the image of a gauge-fixed completion operator subject to boundary and source-sector constraints, while the cokernel condition gives a no-go criterion in the chosen ansatz class. We illustrate the framework with a Schwarzschild example in vacuum GR. Using an ONJA-type complexification different from the Kerr-generating one, we obtain a non-Kerr rotating trial metric and compute its obstruction. The example gives $n_\star=2$ and $\ell_\star=0$, with an additional quadrupolar component at the same order, and shows how the leading residual is removed by the minimal even-parity correction restoring the Kerr complexification. The framework replaces the search for the correct complexification rule with computable criteria for success, completion, or obstruction.
\end{abstract}


\maketitle

\section{Introduction} \label{sec1}
The Kerr geometry is the canonical stationary, axisymmetric, asymptotically flat vacuum black-hole solution and supplies the background by which most relativistic signatures of astrophysical rotation are modeled \cite{Kerr:1963ud}. Its derivation from the Schwarzschild metric through the Newman--Janis algorithm, followed by the analogous construction of Kerr--Newman from the Reissner--Nordstr\"om seed, explains the enduring appeal of the method: it appears to generate rotation without solving the full stationary and axisymmetric field equations \cite{Newman:1965tw,Newman:1965my}.

In order to avoid ambiguity, we distinguish throughout this work between the original Newman--Janis algorithm and modified Newman--Janis prescriptions. We use \emph{ONJA} to denote the original complex-coordinate prescription introduced by Newman and Janis, in which the radial coordinate is allowed to take complex values and functions of \(r\) are then promoted to real functions of the complexified variables. We use \emph{MNJA} to denote modified procedures, including non-complexification variants, gauge-field extensions, Giampieri-type simplifications, and hybrid prescriptions designed to reduce or replace the ambiguous complexification step. Finally, we use \emph{Newman--Janis-type deformation} as an umbrella term for the rotating trial configuration produced by either ONJA or MNJA. The obstruction-completion framework developed below applies to this broader class. It is not a claim that the ONJA, by itself, is a universal solution-generating theorem.

The difficulty is that these successes do not make the Newman--Janis algorithm a general solution-generating theorem. Its steps are not known to follow from a variational principle, a symmetry of the field equations, a B\"acklund transformation, or a uniqueness theorem for rotating deformations of a static seed. A B\"acklund transformation is a map that takes one exact solution of a differential system into another exact solution, usually through an auxiliary set of first-order relations. We use the term only to emphasize that the Newman--Janis algorithm is not known to arise from such a solution-generating symmetry of the field equations in the general settings considered here. Drake and Szekeres showed that the Kerr--Newman case can be sharpened by reducing part of the original ambiguity, but their analysis also indicates how constrained this success is \cite{Drake:1998gf}. Thus, outside special settings, the algorithm should be treated as an ansatz-generating prescription whose dynamical validity must be checked.

The structural problem is not merely the use of complex coordinates, which can be meaningful in special settings such as Kerr--Schild families \cite{Gurses:1975vu}. Rather, a static seed fixes only the nonrotating monopolar geometry and its source interpretation; it does not determine the frame-dragging sector, quadrupolar deformation, angular stress anisotropies, matter-field deformation, or hidden-symmetry structure of a rotating spacetime. This nonuniqueness is visible in regular black holes, nonlinear electrodynamics, and modified-gravity applications, where Newman--Janis outputs may fail the intended equations, develop pathological effective sources, or no longer represent the original matter model \cite{Bambi:2013ufa,Toshmatov:2017zpr,Hansen:2013owa,CiriloLombardo:2004qw}.

Existing strategies for assessing Newman--Janis outputs are indispensable but largely operate as post-construction consistency checks. One may substitute the resulting metric into the gravitational field equations, test whether a Boyer--Lindquist-type coordinate system exists, check circularity, inspect curvature invariants, study separability of geodesic or wave equations, or reconstruct an effective stress tensor. Modified Newman--Janis prescriptions, including non-complexification variants, gauge-field extensions, and hybrid procedures combining non-complexification with additional transformations, reduce some of the technical ambiguity in particular classes of examples \cite{Azreg-Ainou:2014pra,Erbin:2014aya,Erbin:2016lzq,Chaturvedi:2023ctn}. Reviews and earlier applications also show that the algorithm has been used in several gravitational and matter sectors, with varying degrees of success \cite{Canonico:2011lba,Demianski:1966kerrnut}. These developments are valuable, but they do not by themselves convert either ONJA or MNJA into a general dynamical map. They typically begin after a rotating metric has already been guessed.

In this paper, we instead treat a Newman--Janis deformation as an off-shell map. Its output is not assumed to solve the field equations; its failure is encoded in a residual tensor obtained by substituting the trial geometry and matter fields into the intended dynamical system. The vanishing of this tensor is the special case in which the prescription succeeds. Otherwise, the residual defines an obstruction and shifts the question from which complexification is correct to whether the obstruction lies in the image of an appropriate residual-completion operator. If it does, the ansatz can be corrected perturbatively within a specified metric and matter sector; if it does not, the result is a no-go statement within the chosen ansatz class, boundary conditions, and source-preservation assumptions.

This viewpoint also organizes familiar Kerr-like consistency checks. Circularity and Boyer--Lindquist integrability are additional geometrical requirements, not consequences of stationarity and axisymmetry alone. Separability is an even stronger hidden-symmetry condition associated with the Kerr family \cite{Carter:1968rr}. Likewise, reconstructing an effective stress tensor does not prove preservation of the seed equation of state, Segre type, or matter-field origin; recent Newman--Janis analyses show that rotating and nonrotating stress tensors may describe different physical systems \cite{Beltracchi:2021ris,Beltracchi:2021tcx}. We therefore incorporate these checks as obstruction channels rather than as disconnected after-the-fact tests.

The aim of this work is to formulate an obstruction and residual-completion theory for Newman--Janis deformations. We construct a dynamical obstruction tensor, decompose it into field-equation, geometrical, coordinate-admissibility, and matter-sector channels, and ask whether a minimal correction of the metric and matter fields can cancel the leading obstruction. The goal is not to show that every failed Newman--Janis metric can be repaired, but to distinguish success, controlled completion, and obstruction in a precise perturbative framework.

The scope of the analysis is deliberately restricted. We assume stationary and axisymmetric deformations of static spherical seeds and work perturbatively in the rotation parameter when residual completion is required. In addition to stationarity and axisymmetry, the rotating candidate is required to satisfy the regularity conditions appropriate to an axial spacetime: smoothness on the fixed-point set of the axial Killing field, absence of conical defects on the rotation axis, and invariance under reflection through the equatorial plane. These requirements are not consequences of the Newman--Janis prescription itself and are therefore treated as independent geometrical admissibility conditions. We impose gauge and minimality conditions so that the completion problem is not made vacuous by arbitrary deformations. We also require source preservation when the physical interpretation of the seed is part of the problem. A metric that solves Einstein's equations with some effective stress tensor is not automatically an acceptable rotating counterpart of a nonlinear-electrodynamic, anisotropic-fluid, scalar-field, or modified-gravity seed. The framework developed here is designed to keep these distinctions explicit.

The paper is organized as follows. Section \ref{sec2} defines static seeds and off-shell Newman--Janis-type maps. Section \ref{sec3} introduces the obstruction tensor and its perturbative expansion. Section \ref{sec4} treats circularity, Boyer--Lindquist integrability, axis regularity, and reflection symmetry. Section \ref{sec5} develops residual completion with gauge, boundary, and minimality conditions. Section \ref{sec6} analyzes source-preservation obstructions, while Sec. \ref{sec7} states the obstruction and no-go criteria. Section \ref{sec8} gives a concrete Schwarzschild example in which a non-Kerr ONJA complexification produces an explicit obstruction with \(n_\star=2\) and \(\ell_\star=0\). Section \ref{sec9} summarizes the framework and outlines extensions.

We use the metric signature $(-,+,+,+)$ and geometrized units by setting $G=c=1$. Greek indices denote spacetime components. All covariant derivatives and curvature tensors are those of the Levi--Civita connection of the metric under consideration. Unless explicitly stated otherwise, the rotation parameter is treated as a deformation parameter, and the nonrotating seed is recovered smoothly in the zero-rotation limit.

\section{Static Seeds and Newman--Janis-Type Deformations} \label{sec2}
\subsection{Stationarity, axisymmetry, and circularity}
We now isolate the purely geometrical restrictions that a Newman--Janis-type output must satisfy if it is to be interpreted as a Kerr-like rotating geometry. Let \((\mathcal{M},g_{\mu\nu})\) be a four-dimensional spacetime admitting a stationary Killing field \(\xi^\mu\) and an axial Killing field \(\eta^\mu\). These vector fields obey
\begin{equation}
\mathcal{L}_{\xi}g_{\mu\nu}
=
0,
\quad
\mathcal{L}_{\eta}g_{\mu\nu}
=
0,
\label{4.1}
\end{equation}
or equivalently
\begin{equation}
\nabla_\mu \xi_\nu+\nabla_\nu\xi_\mu
=
0,
\quad
\nabla_\mu \eta_\nu+\nabla_\nu\eta_\mu
=
0 .
\label{4.2}
\end{equation}
The axial Killing field is assumed to have closed spacelike orbits away from the rotation axis. For the stationary and axial symmetries relevant to rotating black holes, we also assume that the two Killing fields commute,
\begin{equation}
[\xi,\eta]^\mu
=
\xi^\nu\nabla_\nu\eta^\mu
-
\eta^\nu\nabla_\nu\xi^\mu
=
0 .
\label{4.3}
\end{equation}
Equations \eqref{4.1}--\eqref{4.3} are standard in the geometric formulation of stationary and axisymmetric spacetimes \cite{Wald:1984rg,Carter_1970,Stephani:2003tm}.

In a coordinate system adapted to the Killing fields,
\begin{equation}
\xi^\mu
=
\left(\frac{\partial}{\partial t}\right)^\mu,
\quad
\eta^\mu
=
\left(\frac{\partial}{\partial \phi}\right)^\mu ,
\label{4.4}
\end{equation}
the metric components are independent of \(t\) and \(\phi\). Stationarity and axisymmetry alone, however, do not force the metric to have the familiar circular form. A general stationary and axisymmetric metric may contain cross terms between the Killing directions \((t,\phi)\) and the meridional directions \((r,\theta)\). In adapted coordinates one may write schematically
\begin{equation}
ds^2
=
\gamma_{AB}(x^c)\,dy^A dy^B
+
2q_{Aa}(x^c)\,dy^A dx^a
+
h_{ab}(x^c)\,dx^a dx^b,
\label{4.5}
\end{equation}
where \(y^A=(t,\phi)\), \(x^a=(r,\theta)\), and all metric functions depend only on \(r\) and \(\theta\). The spacetime is circular if the two-dimensional distribution orthogonal to the Killing orbits is integrable. Equivalently, there must exist two-dimensional surfaces whose tangent vectors are everywhere orthogonal to both \(\xi^\mu\) and \(\eta^\mu\).

By Frobenius' theorem, this orthogonal transitivity condition is equivalent to
\begin{equation}
\xi\wedge \eta\wedge d\xi
=
0,
\quad
\xi\wedge \eta\wedge d\eta
=
0 .
\label{4.6}
\end{equation}
In component form, Eq. \eqref{4.6} becomes
\begin{equation}
\xi_{[\mu}\eta_\nu\nabla_\rho\xi_{\sigma]}
=
0,
\quad
\xi_{[\mu}\eta_\nu\nabla_\rho\eta_{\sigma]}
=
0 .
\label{4.7}
\end{equation}
These are the circularity conditions used throughout this paper \cite{Papapetrou:1966zz,Kundt:1966zz,Carter_1970,Stephani:2003tm}. They are local and covariant, and therefore they do not depend on a special radial coordinate or on the existence of a Boyer--Lindquist chart.

If Eq. \eqref{4.7} holds, one may choose coordinates on the meridional two-surfaces so that the metric takes the block-circular form
\begin{align}
ds^2
&=
g_{tt}(r,\theta)\,dt^2
+
2g_{t\phi}(r,\theta)\,dt\,d\phi
+
g_{\phi\phi}(r,\theta)\,d\phi^2
\nn &+
g_{rr}(r,\theta)\,dr^2
+
g_{\theta\theta}(r,\theta)\,d\theta^2
+
2g_{r\theta}(r,\theta)\,dr\,d\theta .
\label{4.8}
\end{align}
With an additional coordinate choice on the two-dimensional meridional space, the \(dr\,d\theta\) term may often be removed locally. The important point is that Eq. \eqref{4.8} is not a consequence of stationarity and axisymmetry alone. It is a consequence of the stronger circularity condition.

In vacuum general relativity and in electrovacuum under the usual symmetry assumptions on the electromagnetic field, circularity follows from the field equations and appropriate regularity assumptions. This is the content of the classical circularity theorems associated with Papapetrou, Kundt--Tr\"umper, and Carter \cite{Papapetrou:1966zz,Kundt:1966zz,Carter_1970}. For generic matter fields, nonlinear electrodynamics, scalar sectors, fluids with meridional flow, or modified gravity, circularity need not follow automatically. Therefore, a Newman--Janis-generated rotating metric must be tested for circularity unless the relevant circularity theorem has already been established for the theory and matter sector under consideration.

The circularity conditions can also be expressed through the Ricci tensor. For a stationary and axisymmetric solution of Einstein's equations, sufficient field-equation conditions for circularity take the schematic form
\begin{equation}
\xi^\alpha R_{\alpha[\mu}\xi_\nu\eta_{\rho]}
=
0,
\quad
\eta^\alpha R_{\alpha[\mu}\xi_\nu\eta_{\rho]}
=
0 ,
\label{4.9}
\end{equation}
together with appropriate boundary or axis regularity conditions. Using Einstein's equations, Eq. \eqref{4.9} becomes a condition on the stress tensor. Thus, circularity is partly geometrical and partly dynamical: in special systems it is enforced by the field equations, while in a general off-shell Newman--Janis deformation it must be imposed as an admissibility condition.

For the present framework, we therefore define a geometrically circular Newman--Janis deformation by
\begin{equation}
\boxed{
\xi_{[\mu}\eta_\nu\nabla_\rho\xi_{\sigma]}
=
0,
\quad
\xi_{[\mu}\eta_\nu\nabla_\rho\eta_{\sigma]}
=
0 .
}
\label{4.10}
\end{equation}
If either of these tensors is nonzero, the Newman--Janis output is non-circular. Such a metric may still be a legitimate stationary and axisymmetric geometry, but it is not Kerr-like in the circular sense and it cannot be reduced to the usual Boyer--Lindquist block form without further conditions.

\subsection{Static spherical seed geometries}
We begin with a static, spherically symmetric seed spacetime written in a Schwarzschild-type radial gauge. The metric is assumed to admit a hypersurface-orthogonal timelike Killing field and an \(SO(3)\) action with spacelike two-sphere orbits. In the coordinate system \((t,r,\theta,\phi)\), the seed line element is taken to be
\begin{equation}
ds_{\rm stat}^2
=
-F(r)\,dt^2
+
\frac{dr^2}{F(r)}
+
R^2(r)\,d\Omega_2^2,
\label{2.1}
\end{equation}
where $d\Omega_2^2=d\theta^2+\sin^2\theta\,d\phi^2 .$
The special case \(R(r)=r\) is the usual areal-radius gauge. We keep \(R(r)\) explicit in order to separate two logically distinct assumptions: the existence of a static spherical seed and the additional choice that the radial coordinate coincides with the areal radius. The form \eqref{2.1} is not meant to represent the most general static spherical geometry in an arbitrary radial coordinate. It is the seed class relevant for the Newman--Janis-type constructions considered in this work.

Following the standard retarded Eddington--Finkelstein construction, we introduce the null coordinate
\begin{equation}
u
=
t-r_\ast(r),
\quad
\frac{dr_\ast}{dr}
=
\frac{1}{F(r)} .
\label{2.2}
\end{equation}
Substituting \(dt=du+F^{-1}(r)dr\) into Eq. \eqref{2.1} gives
\begin{equation}
ds_0^2
=
-F(r)\,du^2
-2\,du\,dr
+
R^2(r)
\left(
d\theta^2+\sin^2\theta\,d\phi^2
\right).
\label{2.3}
\end{equation}
Equations \eqref{2.2} and \eqref{2.3} are the sign convention used here for the outgoing Eddington--Finkelstein form. The corresponding ingoing convention is obtained by reversing the sign of the \(du\,dr\) term. This horizon-penetrating coordinate form is the natural starting point for the tetrad version of the Newman--Janis algorithm, where the inverse metric rather than the covariant metric is first complexified \cite{Carroll_2019,Erbin:2014aya}.

The nonzero inverse metric components following from Eq. \eqref{2.3} are
\begin{align}
&g_{(0)}^{ur}
=
g_{(0)}^{ru}
=
-1,
\quad
g_{(0)}^{rr}
=
F(r),
\nn
&g_{(0)}^{\theta\theta}
=
\frac{1}{R^2(r)},
\quad
g_{(0)}^{\phi\phi}
=
\frac{1}{R^2(r)\sin^2\theta}.
\label{2.4}
\end{align}
The pair \((F,R)\), together with the seed matter fields, will be treated as the seed input. We denote the matter fields collectively by \(\Psi^{(0)}\). The seed is assumed to satisfy the equations of motion of a specified gravitational system,
\begin{align}
\mathfrak{E}_{\mu\nu}
\left[g^{(0)},\Psi^{(0)}\right]
\equiv
&\mathcal{E}_{\mu\nu}
\left[g^{(0)},\Psi^{(0)}\right]
-
8\pi
T_{\mu\nu}
\left[g^{(0)},\Psi^{(0)}\right]
=
0,
\nn
&\mathfrak{S}_I
\left[g^{(0)},\Psi^{(0)}\right]
=
0 .
\label{2.5}
\end{align}
Here \(\mathcal{E}_{\mu\nu}\) denotes the metric field-equation tensor of the chosen theory, \(T_{\mu\nu}\) is the stress tensor of the specified matter sector, and \(\mathfrak{S}_I=0\) denotes the remaining matter equations. In Einstein gravity, \(\mathcal{E}_{\mu\nu}=G_{\mu\nu}\), but the notation is deliberately broader because the obstruction-completion framework is intended to apply also to nonlinear electrodynamics, scalar-tensor models, and effective modified-gravity systems.

Static spherical symmetry restricts the seed stress tensor, in an orthonormal frame adapted to the static observers, to the algebraic form
\begin{equation}
T_{\hat a\hat b}^{(0)}
=
{\rm diag}
\left(
\rho_0,
p_{r0},
p_{\perp0},
p_{\perp0}
\right).
\label{2.6}
\end{equation}
The equality of the two angular pressures follows from the \(SO(3)\) symmetry. A rotating deformation need not preserve this eigenvalue structure. This observation is one of the reasons why a Newman--Janis output must not be identified automatically with a rotating solution of the same matter model.

\subsection{Original, modified, and off-shell Newman--Janis-type deformation maps}
For the seed metric \eqref{2.3}, a convenient null tetrad is
\begin{align}
\ell^\mu
=
\delta^\mu_r,
\quad
n^\mu
=
\delta^\mu_u
-
\frac{F(r)}{2}\delta^\mu_r,
\nn
m^\mu
=
\frac{1}{\sqrt{2}R(r)}
\left(
\delta^\mu_\theta
+
\frac{i}{\sin\theta}\delta^\mu_\phi
\right),
\label{2.7}
\end{align}
with \(\bar m^\mu\) the complex conjugate of \(m^\mu\). In the mostly-plus convention, these vectors obey
\begin{equation}
\ell^\mu \ell_\mu
=
n^\mu n_\mu
=
m^\mu m_\mu
=
0,
\quad
\ell^\mu n_\mu
=
-1,
\quad
m^\mu \bar m_\mu
=
1,
\label{2.8}
\end{equation}
and reconstruct the inverse metric as
\begin{equation}
g_{(0)}^{\mu\nu}
=
-\ell^\mu n^\nu
-\ell^\nu n^\mu
+
m^\mu\bar m^\nu
+
m^\nu\bar m^\mu .
\label{2.9}
\end{equation}
The use of a null tetrad and the reconstruction of the inverse metric are part of the standard Newman--Janis prescription \cite{Newman:1965tw,Erbin:2014aya}.

We now separate the original Newman--Janis algorithm from its modified variants. In the original Newman--Janis algorithm (ONJA), the coordinates \(u\) and \(r\) are complexified and then transformed according to a rule of the representative form
\begin{equation}
u
=
u'
+
ia\cos\theta,
\quad
r
=
r'
-
ia\cos\theta,
\quad
\theta'
=
\theta,
\quad
\phi'
=
\phi ,
\label{2.10}
\end{equation}
after which \(u'\) and \(r'\) are again treated as real coordinates. The parameter \(a\) is interpreted as the rotation parameter. Equation \eqref{2.10} is the standard complex coordinate step in the tetrad formulation of the ONJA \cite{Newman:1965tw,Newman:1965my,Drake:1998gf,Canonico:2011lba}. The nontrivial ambiguity is not the formal writing of Eq. \eqref{2.10}, but the rule by which functions of \(r\) are promoted to real functions of the complexified variables. In the ONJA this is expressed schematically as
\begin{align}
F(r)
\longmapsto
\widetilde F(r,\bar r;\theta,a),
\quad
R^2(r)
\longmapsto
\widetilde\Sigma(r,\bar r;\theta,a),
\nn
\widetilde F(r,r;\theta,0)
=
F(r),
\quad
\widetilde\Sigma(r,r;\theta,0)
=
R^2(r).
\label{2.11}
\end{align}
Equation \eqref{2.11} should therefore be read as the complexification step of the ONJA, not as a statement about all modified Newman--Janis prescriptions. The complexification rules for \(F(r)\) and \(R^2(r)\) may be different; in particular, the rule used to promote the lapse function need not coincide with the rule used to promote the areal factor.

Operationally, the ONJA does not complexify the covariant metric components directly. One first writes the inverse seed metric in terms of the null tetrad, promotes the radial functions appearing in the tetrad to real functions of the complexified variables according to the chosen rule, applies the complex coordinate transformation to the tetrad vectors, and finally reconstructs a real inverse metric using
\begin{equation}
g^{\mu\nu}
=
-\ell^\mu n^\nu
-\ell^\nu n^\mu
+
m^\mu\bar m^\nu
+
m^\nu\bar m^\mu .
\label{eq:onja_inverse_metric_reconstruction}
\end{equation}
The covariant rotating metric is then obtained by matrix inversion. This is why different complexification rules for the same static seed can lead to different rotating trial metrics. The explicit Schwarzschild example in Sec. \ref{sec8} illustrates this procedure.

Modified Newman--Janis algorithms (MNJA) alter this step. In non-complexification variants, one does not complexify the radial functions in the ONJA sense. Instead, the complex coordinate transformation is either replaced or supplemented by real deformation functions chosen so that the rotating ansatz has the desired static limit, coordinate structure, and, in favorable cases, field-equation closure \cite{Azreg-Ainou:2014pra,Erbin:2014aya,Erbin:2016lzq}. Other modified prescriptions include gauge-field extensions and hybrid procedures. For example, recent work on constant-curvature \(f(R)\) gravity combines non-complexification of the radial coordinate with an additional prescription in the angular sector to obtain exact rotating solutions in the corresponding theory \cite{Chaturvedi:2023ctn}. Earlier Newman--Janis-related constructions, including the Kerr--NUT solution, also illustrate that the complexification idea belongs to a broader family of solution-generating or ansatz-generating procedures rather than to a unique universal algorithm \cite{Demianski:1966kerrnut}.

Thus, the present paper does not assume that the ONJA is the unique or preferred prescription. Different ONJA and MNJA choices may agree in the Schwarzschild--Kerr and Reissner--Nordstr\"om--Kerr--Newman cases while differing for more general seeds. This is precisely where the algorithm ceases to be a theorem and becomes an ansatz-generating rule. The obstruction framework below is designed to test the output of a specified prescription, whether that output was obtained by ONJA, MNJA, or another Newman--Janis-type construction.

We therefore define a Newman--Janis-type deformation in this work as an off-shell map associated with a specified prescription \(P\), where \(P\) may be ONJA, MNJA, or another explicitly stated Newman--Janis-type ansatz rule:
\begin{align}
\mathfrak{N}^{P}_a&:\left(
g_{\mu\nu}^{(0)},\Psi^{(0)};
\mathcal{E}_{\mu\nu},T_{\mu\nu},\mathfrak{S}_I
\right)
\longmapsto
\left(
g_{\mu\nu}^{P}(a),
\Psi^{P,{\rm trial}}(a)
\right),
\nn
&P\in\{\mathrm{ONJA},\mathrm{MNJA},\mathrm{NJA\mbox{-}type}\}.
\label{2.12}
\end{align}
This definition differs from the usual informal use of the algorithm in one essential respect: the image of \(\mathfrak{N}^{P}_a\) is not assumed to solve the field equations. It is only a rotating trial configuration associated with a seed, a theory, a matter model, and a specified ONJA or MNJA prescription. When no distinction is needed, we write \(g_{\mu\nu}^{\rm NJA}\) as a shorthand for this Newman--Janis-type trial metric. This shorthand does not mean that the ONJA has been used unless the prescription is explicitly labelled \(P=\mathrm{ONJA}\). The map must satisfy the zero-rotation consistency conditions
\begin{equation}
\lim_{a\to0}
g_{\mu\nu}^{P}(a)
=
g_{\mu\nu}^{(0)},
\quad
\lim_{a\to0}
\Psi^{P,{\rm trial}}(a)
=
\Psi^{(0)} .
\label{2.13}
\end{equation}
No condition of dynamical admissibility is imposed at this stage. In particular, we do not assume that
\begin{equation}
\mathfrak{E}_{\mu\nu}
\left[
g^{P}(a),
\Psi^{P,{\rm trial}}(a)
\right]
=
0,
\quad
\mathfrak{S}_I
\left[
g^{P}(a),
\Psi^{P,{\rm trial}}(a)
\right]
=
0 .
\label{2.14}
\end{equation}
The purpose of Eqs. \eqref{2.12}--\eqref{2.14} is to separate four operations that are often conflated: choosing an ONJA or MNJA prescription, constructing the corresponding rotating trial metric, assigning trial matter fields, and proving that the resulting configuration solves a specified dynamical system.

For matter fields, Eq. \eqref{2.12} should be understood with particular care. In the Einstein--Maxwell case, the gauge potential can be transformed consistently only after a suitable gauge choice, and the transformation of the gauge field is not automatically fixed by the metric transformation alone \cite{Erbin:2014aya}. For nonlinear electrodynamics, scalar fields, anisotropic fluids, or modified-gravity auxiliary fields, the ambiguity is stronger. The notation \(\Psi^{\rm trial}(a)\) is therefore intentional: it denotes the matter configuration suggested by a Newman--Janis-type rule, not yet a verified rotating matter solution.

\subsection{Minimal rotating ansatz and deformation functions}
The off-shell viewpoint allows us to replace a single complexification rule by a controlled stationary and axisymmetric ansatz. We use coordinates \((u,r,\theta,\phi)\) adapted to the two commuting vector fields \(\partial_u\) and \(\partial_\phi\). Reflection symmetry across the equatorial plane and reversal of the rotation parameter restrict the parity of the metric components. A minimal Eddington--Finkelstein-type rotating metric compatible with these symmetries is
\begin{equation}
\begin{aligned}
&ds_{\rm off}^2
={}
-A(r,\theta;a)\,du^2
-2B(r,\theta;a)\,du\,dr
\nn&+2a\sin^2\theta\,C(r,\theta;a)\,du\,d\phi
+
2a\sin^2\theta\,D(r,\theta;a)\,dr\,d\phi
\nn&+
E(r,\theta;a)\,d\theta^2
+
\sin^2\theta\,G(r,\theta;a)\,d\phi^2 .
\end{aligned}
\label{2.15}
\end{equation}
This ansatz is not yet a solution ansatz in the dynamical sense. It is a kinematic container large enough to include the usual Newman--Janis output, while retaining enough structure to define a meaningful completion problem.

The ansatz \eqref{2.15} is also subject to three regularity and symmetry requirements that will be used later as admissibility tests. First, the axial Killing field \(\eta^\mu=(\partial/\partial\phi)^\mu\) must have a regular fixed-point set at \(\theta=0,\pi\). Since \(\eta^2=g_{\phi\phi}\), the axis is characterized by \(g_{\phi\phi}=0\). Smoothness near the axis requires the mixed axial components to vanish at least as
\begin{equation}
g_{\phi A}=O(\sin^2\theta),
\quad
A\neq \phi,
\quad
\theta\to0,\pi ,
\label{eq:axis_smoothness_conditions}
\end{equation}
with all coefficient functions finite in a local Cartesian chart transverse to the axis. The factors of \(\sin^2\theta\) in \(g_{u\phi}\), \(g_{r\phi}\), and \(g_{\phi\phi}\) in Eq. \eqref{2.15} are chosen precisely to make this possible, but finiteness of the functions \(C,D,G\) and the elementary-flatness condition below must still be imposed.

Second, the axis must be free of a conical deficit or excess. The invariant elementary-flatness condition is
\begin{equation}
\lim_{\eta^2\to0}
\frac{
\nabla_\mu(\eta^2)\nabla^\mu(\eta^2)
}
{4\eta^2}
=
1,
\quad
\eta^2=g_{\phi\phi}.
\label{eq:elementary_flatness_invariant}
\end{equation}
For the coordinate form \eqref{2.15}, this becomes
\begin{equation}
\lim_{\theta\to0,\pi}
\frac{g_{\phi\phi}}
{g_{\theta\theta}\sin^2\theta}
=
1.
\label{eq:elementary_flatness_coordinate}
\end{equation}
Equivalently, since \(g_{\theta\theta}=E(r,\theta;a)\) and \(g_{\phi\phi}=\sin^2\theta\,G(r,\theta;a)\), one must impose
\begin{equation}
G(r,0;a)=E(r,0;a),
\quad
G(r,\pi;a)=E(r,\pi;a).
\label{eq:no_conical_defect_EG}
\end{equation}
If Eq. \eqref{eq:no_conical_defect_EG} fails, the rotating candidate contains a conical singularity on the corresponding segment of the symmetry axis.

Third, equatorial reflection symmetry is imposed as invariance under
\begin{equation}
\mathcal{R}_{\rm eq}:\theta\mapsto \pi-\theta .
\label{eq:equatorial_reflection_map}
\end{equation}
For the reduced ansatz \eqref{2.15}, this requires the scalar coefficient functions to obey
\begin{equation}
X(r,\pi-\theta;a)=X(r,\theta;a),
\quad
X\in\{A,B,C,D,E,G\}.
\label{eq:equatorial_reflection_functions}
\end{equation}
Equivalently, these functions are smooth functions of \(\cos^2\theta\) near the equatorial plane. In a more general stationary and axisymmetric ansatz, tensor components carrying one \(\theta\)-index would be odd under \(\theta\mapsto\pi-\theta\), while components carrying no \(\theta\)-index or two \(\theta\)-indices would be even. Thus equatorial reflection symmetry is not merely a notation convention; it is a boundary and parity condition on the admissible rotating deformation.

The zero-rotation limit fixes the boundary values
\begin{align}
A(r,\theta;0)
=
F(r),
\quad
B(r,\theta;0)
=
1,
\nn
E(r,\theta;0)
=
R^2(r),
\quad
G(r,\theta;0)
=
R^2(r).
\label{2.16}
\end{align}
The functions \(C\) and \(D\) need only remain finite as \(a\to0\), since their contribution to the metric in Eq. \eqref{2.15} is already multiplied by \(a\). The leading frame-dragging sector therefore appears in the \(a\,du\,d\phi\) and \(a\,dr\,d\phi\) components, while the leading quadrupolar deformations of the diagonal metric components begin at order \(a^2\).

For a parity-symmetric rotating configuration, the metric functions may be expanded as
\begin{equation}
\begin{aligned}
A(r,\theta;a)
&=
F(r)
+
a^2 A_2(r,\theta)
+
O(a^4),
\\
B(r,\theta;a)
&=
1
+
a^2 B_2(r,\theta)
+
O(a^4),
\\
C(r,\theta;a)
&=
C_0(r,\theta)
+
a^2 C_2(r,\theta)
+
O(a^4),
\\
D(r,\theta;a)
&=
D_0(r,\theta)
+
a^2 D_2(r,\theta)
+
O(a^4),
\\
E(r,\theta;a)
&=
R^2(r)
+
a^2 E_2(r,\theta)
+
O(a^4),
\\
G(r,\theta;a)
&=
R^2(r)
+
a^2 G_2(r,\theta)
+
O(a^4).
\end{aligned}
\label{2.17}
\end{equation}
The odd powers of \(a\) appear in the metric only through the axial cross terms in Eq. \eqref{2.15}. This is the tensorial expression of the fact that reversing the sense of rotation should reverse frame dragging but should not change the even-parity quadrupolar deformation.

The usual Kerr-like angular combination is recovered from the Newman--Janis construction by the replacement
\begin{equation}
\Sigma(r,\theta;a)
=
R^2(r)
+
a^2\cos^2\theta .
\label{2.18}
\end{equation}
For \(R(r)=r\), Eq. \eqref{2.18} becomes the standard Kerr expression \(\Sigma=r^2+a^2\cos^2\theta\), which appears in the Kerr and Kerr--Newman geometries and in many Newman--Janis-generated rotating regular black-hole metrics \cite{Kerr:1963ud,Newman:1965my,Azreg-Ainou:2014pra}. In the present framework, however, Eq. \eqref{2.18} is not imposed as a theorem. It is a Kerr-inspired angular structure that may or may not be compatible with the field equations and matter-sector constraints of a chosen seed.

We now separate the raw Newman--Janis output from its controlled enlargement. Let
\begin{equation}
g_{\mu\nu}^{\rm off}(a)
=
g_{\mu\nu}^{\rm NJA}(a)
+
h_{\mu\nu}
\left[
\mathcal{F}_I
\right](a),
\quad
\lim_{a\to0}
h_{\mu\nu}
\left[
\mathcal{F}_I
\right](a)
=
0 ,
\label{2.19}
\end{equation}
where the finite collection \(\mathcal{F}_I\) denotes deformation functions chosen within the stationary, axisymmetric, reflection-symmetric ansatz class. Similarly, for the matter sector we write
\begin{equation}
\Psi^{\rm off}(a)
=
\Psi^{\rm trial}(a)
+
\Xi
\left[
\mathcal{P}_J
\right](a),
\quad
\lim_{a\to0}
\Xi
\left[
\mathcal{P}_J
\right](a)
=
0 .
\label{2.20}
\end{equation}
The functions \(\mathcal{P}_J\) represent matter-sector corrections, such as corrections to an electromagnetic potential, scalar profile, anisotropic-fluid tetrad, or nonlinear-electrodynamic invariant.

Equations \eqref{2.19} and \eqref{2.20} define the minimal off-shell completion space. The word minimal does not mean unique. It means that the ansatz contains only those deformation sectors required by stationarity, axisymmetry, reflection symmetry, the static limit, and the anticipated leading obstruction channel. In later sections, the functions \(\mathcal{F}_I\) and \(\mathcal{P}_J\) will be fixed, constrained, or ruled out by the field equations, geometric integrability conditions, and source-preservation requirements.

\section{The Newman--Janis Obstruction Tensor} \label{sec3}
\subsection{Definition of the obstruction tensor}
The off-shell viewpoint introduced in Sec. \ref{sec2} allows the failure of a Newman--Janis-type deformation to be represented by a tensorial residual. Let the metric and matter equations of the target theory be written as
\begin{equation}
\mathfrak{E}_{\mu\nu}
\left[g,\Psi\right]
\equiv
\mathcal{E}_{\mu\nu}
\left[g,\Psi\right]
-
8\pi
T_{\mu\nu}
\left[g,\Psi\right]
=
0,
\quad
\mathfrak{S}_I
\left[g,\Psi\right]
=
0 .
\label{3.1}
\end{equation}
For Einstein gravity minimally coupled to matter, \(\mathcal{E}_{\mu\nu}=G_{\mu\nu}\), with
\begin{equation}
G_{\mu\nu}
=
R_{\mu\nu}
-
\frac{1}{2}R g_{\mu\nu}.
\label{3.2}
\end{equation}
Equation \eqref{3.2} is the standard definition of the Einstein tensor \cite{Wald:1984rg,Carroll_2019}. The notation in Eq. \eqref{3.1}, however, is deliberately more general. It permits \(\mathcal{E}_{\mu\nu}\) to include cosmological, higher-curvature, scalar-tensor, or effective modified-gravity contributions, provided the chosen theory admits a well-defined symmetric metric equation.

Given an off-shell Newman--Janis-type image generated by a specified prescription \(P\),
\begin{equation}
\mathfrak{N}^{P}_a:
\left(
g_{\mu\nu}^{(0)},\Psi^{(0)}
\right)
\longmapsto
\left(
g_{\mu\nu}^{P}(a),
\Psi^{P,{\rm trial}}(a)
\right),
\label{3.3}
\end{equation}
where \(P\) may denote ONJA, MNJA, or another explicitly stated Newman--Janis-type ansatz rule, we define the corresponding obstruction tensor by
\begin{align}
&\mathcal{O}_{\mu\nu}^{P}(a)
\equiv
\mathfrak{E}_{\mu\nu}
\left[
g^{P}(a),
\Psi^{P,{\rm trial}}(a)
\right]
\nn&=
\mathcal{E}_{\mu\nu}
\left[
g^{P}(a),
\Psi^{P,{\rm trial}}(a)
\right]
-
8\pi
T_{\mu\nu}
\left[
g^{P}(a),
\Psi^{P,{\rm trial}}(a)
\right].
\label{3.4}
\end{align}
When the prescription \(P\) is not central to the discussion, we suppress it and write \(\mathcal{O}_{\mu\nu}^{\rm NJA}\), with the understanding that this denotes the obstruction of a specified Newman--Janis-type trial output, not necessarily the output of the ONJA. This is the primary object of the paper. It is not a new field equation. It is the tensorial residue left after the off-shell Newman--Janis output has been substituted into the field equations of the intended theory. The ordinary claim that the Newman--Janis algorithm has generated a solution is recovered only in the special case
\begin{equation}
\mathcal{O}_{\mu\nu}^{\rm NJA}(a)
=
0,
\quad
\mathfrak{S}_I
\left[
g^{\rm NJA}(a),
\Psi^{\rm trial}(a)
\right]
=
0 .
\label{3.5}
\end{equation}
Thus, Eq. \eqref{3.5} is the dynamical admissibility condition for the uncorrected Newman--Janis output.

For pure Einstein gravity, the obstruction tensor reduces to
\begin{equation}
\mathcal{O}_{\mu\nu}^{\rm NJA}
=
G_{\mu\nu}
\left[
g^{\rm NJA}
\right]
-
8\pi
T_{\mu\nu}
\left[
g^{\rm NJA},\Psi^{\rm trial}
\right].
\label{3.6}
\end{equation}
If no independent matter model is supplied, one may define an effective stress tensor by
\begin{equation}
T_{\mu\nu}^{\rm eff}
=
\frac{1}{8\pi}
G_{\mu\nu}
\left[
g^{\rm NJA}
\right].
\label{3.7}
\end{equation}
Equation \eqref{3.7} is often used in the interpretation of rotating metrics generated from static seeds, including Newman--Janis-generated regular black holes and Kerr--Schild systems \cite{Beltracchi:2021ris,Beltracchi:2021tcx}. In the present paper, however, Eq. \eqref{3.7} is not taken as sufficient for physical admissibility. It only says that the geometry can be read as a solution of Einstein's equation with some effective source. It does not establish that this source has the same field-theoretic origin, equation of state, Segre type, or energy-condition structure as the seed matter.

The obstruction tensor also has a useful compatibility interpretation. In Einstein gravity, the contracted Bianchi identity gives
\begin{equation}
\nabla^\mu G_{\mu\nu}
=
0 .
\label{3.8}
\end{equation}
Equation \eqref{3.8} is a geometric identity following from the differential Bianchi identity \cite{Wald:1984rg,Carroll_2019}. Hence, if the matter equations imply covariant stress-tensor conservation, then an exact solution satisfies
\begin{equation}
\nabla^\mu T_{\mu\nu}
=
0 .
\label{3.9}
\end{equation}
For an off-shell Newman--Janis output, the divergence of Eq. \eqref{3.6} gives
\begin{equation}
\nabla^\mu
\mathcal{O}_{\mu\nu}^{\rm NJA}
=
-8\pi
\nabla^\mu
T_{\mu\nu}
\left[
g^{\rm NJA},\Psi^{\rm trial}
\right],
\label{3.10}
\end{equation}
where the covariant derivative is that of \(g_{\mu\nu}^{\rm NJA}\). Therefore, a nonzero divergence of the obstruction indicates either a failure of the trial matter fields to obey their equations of motion or an inconsistency between the metric and the proposed source. In a diffeomorphism-invariant Lagrangian theory, the analogous statement follows from the Noether identity relating the metric equations and the matter equations \cite{Iyer:1994ys}. This gives the obstruction tensor the status of a covariant consistency test of both dynamical and matter-sector failure.

\subsection{Decomposition into independent obstruction channels}
The obstruction tensor in Eq. \eqref{3.4} contains several logically distinct types of failure. A Newman--Janis output may fail because it does not solve the metric equations, because the accompanying trial matter fields do not solve their own equations, because the resulting effective stress tensor changes physical type, or because the stationary axisymmetric geometry does not satisfy the integrability properties expected of a Kerr-like spacetime. We therefore decompose the obstruction into channels,
\begin{equation}
\mathfrak{O}^{\rm NJA}
=
\left(
\mathcal{O}_{\mu\nu}^{\rm dyn},
\mathcal{O}_{I}^{\rm matter},
\mathcal{O}_{A}^{\rm geom},
\mathcal{O}_{B}^{\rm src}
\right).
\label{3.11}
\end{equation}
Here \(\mathfrak{O}^{\rm NJA}\) denotes the full obstruction tuple rather than a single tensor. The first component is the metric-equation obstruction,
\begin{equation}
\mathcal{O}_{\mu\nu}^{\rm dyn}
\equiv
\mathfrak{E}_{\mu\nu}
\left[
g^{\rm NJA},
\Psi^{\rm trial}
\right].
\label{3.12}
\end{equation}
This component is identical to the tensor defined in Eq. \eqref{3.4}. It vanishes if and only if the off-shell Newman--Janis metric and trial matter fields satisfy the metric equations of the target theory.

The second component is the matter-equation obstruction,
\begin{equation}
\mathcal{O}_{I}^{\rm matter}
\equiv
\mathfrak{S}_{I}
\left[
g^{\rm NJA},
\Psi^{\rm trial}
\right].
\label{3.13}
\end{equation}
For a Maxwell field, \(\mathcal{O}_{I}^{\rm matter}\) contains the residual Maxwell equations. For nonlinear electrodynamics, it contains the residual equations associated with the nonlinear constitutive tensor. For a scalar field, it contains the scalar Euler--Lagrange equation. The metric may satisfy Eq. \eqref{3.12} with an effective stress tensor while Eq. \eqref{3.13} fails for the intended matter field. This is a genuine obstruction to interpreting the metric as the rotating version of the chosen seed system.

The third component collects geometrical and coordinate-admissibility obstructions. To make the notation self-contained, we first introduce the quantities entering this channel. Let \(\epsilon_{\mu\nu\rho\sigma}\) be the volume form associated with \(g_{\mu\nu}\). The scalar duals of the two Frobenius circularity conditions are
\begin{align}
\mathcal{C}_{\xi}
&=
\epsilon^{\mu\nu\rho\sigma}
\xi_\mu
\eta_\nu
\nabla_\rho \xi_\sigma ,
\nn
\mathcal{C}_{\eta}
&=
\epsilon^{\mu\nu\rho\sigma}
\xi_\mu
\eta_\nu
\nabla_\rho \eta_\sigma .
\label{eq:geom_channel_circularity_defs}
\end{align}
Thus \(\mathcal{C}_{\xi}=0=\mathcal{C}_{\eta}\) is the scalar form of the circularity condition.

When the Newman--Janis-type output is written in an Eddington--Finkelstein-type stationary and axisymmetric coordinate system \((u,r,\theta,\phi)\), the Boyer--Lindquist-type transformation is tested by candidate radial shifts \(\lambda\) and \(\chi\). Algebraically, these are
\begin{align}
\lambda(r,\theta)
&=
\frac{
-g_{ur}g_{\phi\phi}
+
g_{u\phi}g_{r\phi}
}
{
g_{uu}g_{\phi\phi}
-
g_{u\phi}^{\,2}
},
\nn
\chi(r,\theta)
&=
\frac{
g_{u\phi}g_{ur}
-
g_{uu}g_{r\phi}
}
{
g_{uu}g_{\phi\phi}
-
g_{u\phi}^{\,2}
}.
\label{eq:geom_channel_bl_shifts}
\end{align}
The corresponding Boyer--Lindquist coordinate-integrability obstructions are
\begin{equation}
\mathcal{B}_{\lambda}
=
\partial_\theta \lambda,
\quad
\mathcal{B}_{\chi}
=
\partial_\theta \chi .
\label{eq:geom_channel_bl_obstructions}
\end{equation}
The absence of a conical defect on the two parts of the axis is measured by
\begin{equation}
\mathcal{A}_{\pm}(r;a)
\equiv
\lim_{\theta\to\theta_\pm}
\left[
\frac{
g_{\phi\phi}
}
{
g_{\theta\theta}\sin^2\theta
}
-
1
\right],
\quad
\theta_+=0,
\quad
\theta_-=\pi .
\label{eq:geom_channel_axis_defects}
\end{equation}
Finally, for a reduced ansatz with coefficient functions \(X\in\{A,B,C,D,E,G\}\), the equatorial-reflection residual is
\begin{equation}
\mathcal{R}_{\rm eq}[X](r,\theta;a)
\equiv
X(r,\pi-\theta;a)-X(r,\theta;a).
\label{eq:geom_channel_reflection_residual}
\end{equation}

With these definitions, the geometrical and coordinate-admissibility channel is
\begin{equation}
\mathcal{O}_{A}^{\rm geom}
\equiv
\left(
\mathcal{C}_{\xi},
\mathcal{C}_{\eta},
\mathcal{B}_{\lambda},
\mathcal{B}_{\chi},
\mathcal{A}_{+},
\mathcal{A}_{-},
\mathcal{R}_{\rm eq},
\mathcal{K}_{\rm sep}
\right).
\label{3.14}
\end{equation}
Here \(\mathcal{K}_{\rm sep}\) denotes any selected separability obstruction, such as the failure of a second-rank Killing tensor or principal-tensor ansatz to satisfy its defining equation. This last entry is optional: separability is not required for a stationary axisymmetric metric to be a solution, but it is a strong Kerr-like property and is useful when the target class is required to possess Kerr-type separability. The role of Eq.~\eqref{3.14} is to prevent invariant geometrical requirements, such as circularity, axis regularity, elementary flatness, and reflection symmetry, from being conflated with coordinate-admissibility requirements, such as the existence of a restricted Boyer--Lindquist transformation, or with the metric field equations themselves.

The fourth component is the source-preservation obstruction,
\begin{equation}
\mathcal{O}_{B}^{\rm src}
\equiv
\left(
\Delta{\rm Segre},
\Delta\mathcal{P},
\Delta\mathcal{I}_{\rm matter}
\right).
\label{3.15}
\end{equation}
Here \(\Delta{\rm Segre}\) measures the change in the algebraic type of the mixed stress tensor, \(\Delta\mathcal{P}\) measures the failure of the intended equation-of-state relation, and \(\Delta\mathcal{I}_{\rm matter}\) measures the failure to preserve the field-theoretic interpretation of the source. The use of stress-tensor eigenvalues, Segre type, and equations of state in the physical interpretation of Newman--Janis rotating systems follows the strategy developed in Refs. \cite{Beltracchi:2021ris,Beltracchi:2021tcx}. In this paper these quantities are not merely descriptive labels; they become obstruction quantities that can be used to decide whether a proposed residual completion preserves the seed matter sector.

A Newman--Janis deformation is called dynamically admissible when
\begin{equation}
\mathcal{O}_{\mu\nu}^{\rm dyn}
=
0,
\quad
\mathcal{O}_{I}^{\rm matter}
=
0 .
\label{3.16}
\end{equation}
It is called geometrically admissible, relative to a chosen Kerr-like target class, when
\begin{equation}
\mathcal{O}_{A}^{\rm geom}
=
0 .
\label{3.17}
\end{equation}
It is called source-preserving when
\begin{equation}
\mathcal{O}_{B}^{\rm src}
=
0 .
\label{3.18}
\end{equation}
The strongest notion of admissibility used in this paper is therefore
\begin{equation}
\boxed{
\mathfrak{O}^{\rm NJA}
=
0 .
}
\label{3.19}
\end{equation}
Equation \eqref{3.19} means that the Newman--Janis output solves the intended metric equations, solves the intended matter equations, satisfies the selected geometric integrability conditions, and preserves the seed source class.

The decomposition in Eq. \eqref{3.11} also clarifies a common ambiguity. If \(\mathcal{O}_{\mu\nu}^{\rm dyn}=0\) after defining \(T_{\mu\nu}^{\rm eff}=G_{\mu\nu}/8\pi\), then the metric is dynamically admissible only in a weak effective-source sense. It may still have
\begin{equation}
\mathcal{O}_{B}^{\rm src}
\neq
0 ,
\label{3.20}
\end{equation}
meaning that the rotating metric is not a source-preserving deformation of the original seed. This distinction is essential for regular black holes, nonlinear electrodynamics, and scalar-field seeds, where the physical content of the source is part of the claimed solution.

\subsection{Perturbative expansion of the obstruction}
The residual-completion problem becomes constructive when the obstruction is organized perturbatively in the rotation parameter. We assume that the off-shell Newman--Janis fields admit expansions of the form
\begin{equation}
g_{\mu\nu}^{\rm NJA}(a)
=
g_{\mu\nu}^{(0)}
+
a\,g_{\mu\nu}^{[1]}
+
a^2 g_{\mu\nu}^{[2]}
+
a^3 g_{\mu\nu}^{[3]}
+
O(a^4),
\label{3.21}
\end{equation}
and
\begin{equation}
\Psi^{\rm trial}(a)
=
\Psi^{(0)}
+
a\,\Psi^{[1]}
+
a^2\Psi^{[2]}
+
a^3\Psi^{[3]}
+
O(a^4).
\label{3.22}
\end{equation}
The expansion in Eq. \eqref{3.21} is the standard slow-rotation logic used in relativistic stellar and black-hole perturbation theory: the frame-dragging sector enters at first order in rotation, while the mass, shape, and quadrupolar deformations enter at second order under equatorial reflection symmetry \cite{Hartle:1967he,Hartle:1968si}. In our use, however, Eqs. \eqref{3.21} and \eqref{3.22} are not assumed to solve the perturbative field equations. They only organize the off-shell Newman--Janis residual.

Substitution into Eq. \eqref{3.4} gives
\begin{equation}
\mathcal{O}_{\mu\nu}^{\rm NJA}(a)
=
\mathcal{O}_{\mu\nu}^{(0)}
+
a\,\mathcal{O}_{\mu\nu}^{(1)}
+
a^2\mathcal{O}_{\mu\nu}^{(2)}
+
a^3\mathcal{O}_{\mu\nu}^{(3)}
+
O(a^4).
\label{3.23}
\end{equation}
Because the seed is assumed to satisfy the target equations, Eq. \eqref{3.1} gives
\begin{equation}
\mathcal{O}_{\mu\nu}^{(0)}
=
0 .
\label{3.24}
\end{equation}
The first nonzero coefficient in Eq. \eqref{3.23} defines the leading obstruction order,
\begin{equation}
\boxed{
n_\star
=
\min
\left\{
n\in\mathbb{N}
:
\mathcal{O}_{\mu\nu}^{(n)}
\neq
0
\right\}.
}
\label{3.25}
\end{equation}
If all coefficients vanish to the order considered, then the Newman--Janis output is admissible to that perturbative order. If \(n_\star\) is finite, the residual-completion problem begins at order \(a^{n_\star}\).

The coefficient \(\mathcal{O}_{\mu\nu}^{(n)}\) is obtained by the expansion
\begin{equation}
\mathcal{O}_{\mu\nu}^{(n)}
=
\frac{1}{n!}
\left.
\frac{\partial^n}{\partial a^n}
\mathfrak{E}_{\mu\nu}
\left[
g^{\rm NJA}(a),
\Psi^{\rm trial}(a)
\right]
\right|_{a=0}.
\label{3.26}
\end{equation}
This formula is formal but useful: it defines the obstruction coefficients without committing to a specific coordinate gauge, complexification prescription, or matter model. In a concrete calculation, Eq. \eqref{3.26} is evaluated by substituting the explicit Newman--Janis metric and trial matter fields into the field equations.

Equatorial reflection symmetry imposes additional structure. Under
\begin{equation}
a\mapsto -a,
\quad
\phi\mapsto -\phi ,
\label{3.27}
\end{equation}
the diagonal and meridional metric components are even in \(a\), while the axial cross terms are odd in \(a\). Thus, for a reflection-symmetric rotating deformation,
\begin{align}
g_{uu},
g_{ur},
g_{rr},
g_{\theta\theta},
g_{\phi\phi}
=
\hbox{even functions of }a,
\nn
g_{u\phi},
g_{r\phi}
=
\hbox{odd functions of }a .
\label{3.28}
\end{align}
Consequently, the leading frame-dragging obstruction generally appears in the odd-parity sector at order \(a\), while the leading quadrupolar obstruction generally appears in the even-parity sector at order \(a^2\). This mirrors the Hartle--Thorne separation between first-order rotational dragging and second-order rotational deformation \cite{Hartle:1967he,Hartle:1968si}.

It is useful to project the obstruction onto angular harmonics compatible with stationarity and axisymmetry. For equatorial symmetry, scalar obstruction components may be expanded as
\begin{equation}
\mathcal{O}_{\mu\nu}^{(n)}(r,\theta)
=
\sum_{\ell\in\mathcal{L}_n}
\mathcal{O}_{\mu\nu}^{(n,\ell)}(r)
P_\ell(\cos\theta),
\label{3.29}
\end{equation}
where \(P_\ell\) are Legendre polynomials and \(\mathcal{L}_n\) is the set of allowed multipoles at order \(a^n\). At first order, the axial sector contains the dipolar frame-dragging channel. At second order, the even sector contains monopolar and quadrupolar channels. For the present paper, the practical importance of Eq. \eqref{3.29} is that it identifies the smallest angular sector in which a residual correction must be introduced.

We therefore define the channel-projected leading obstruction by
\begin{equation}
\mathcal{O}_{\mu\nu}^{\star}
\equiv
\mathcal{O}_{\mu\nu}^{(n_\star,\ell_\star)},
\quad
\ell_\star
=
\min
\left\{
\ell\in\mathcal{L}_{n_\star}
:
\mathcal{O}_{\mu\nu}^{(n_\star,\ell)}
\neq
0
\right\}.
\label{3.30}
\end{equation}
The pair \((n_\star,\ell_\star)\) specifies the first obstruction channel. For example, \((1,1)\) indicates a leading frame-dragging failure, whereas \((2,0)\) or \((2,2)\) indicates a leading even-parity deformation failure. This information will be used in Sec. \ref{sec5} to restrict the residual correction to the smallest tensorial sector capable of cancelling the obstruction.

Finally, the perturbative obstruction must obey the linearized compatibility identities inherited from diffeomorphism invariance. In Einstein gravity, expanding Eq. \eqref{3.10} gives, at the first nonvanishing order,
\begin{equation}
\nabla_{(0)}^\mu
\mathcal{O}_{\mu\nu}^{(n_\star)}
=
-8\pi
\mathcal{D}_{\nu}^{(n_\star)}
\left[
\Psi^{[1]},\ldots,\Psi^{[n_\star]}
\right],
\label{3.31}
\end{equation}
where \(\nabla_{(0)}\) is the covariant derivative of the seed metric and \(\mathcal{D}_{\nu}^{(n_\star)}\) is the corresponding order-\(a^{n_\star}\) matter nonconservation functional. If the matter equations are imposed independently to the same order, then
\begin{equation}
\nabla_{(0)}^\mu
\mathcal{O}_{\mu\nu}^{(n_\star)}
=
0 .
\label{3.32}
\end{equation}
Equation \eqref{3.32} is not an additional field equation; it is a compatibility condition. It will become important when the obstruction is treated as a source term for the residual-completion equation.

\section{Boyer--Lindquist Coordinate Integrability and Axis Regularity} \label{sec4}
\subsection{Boyer--Lindquist integrability}
The Newman--Janis algorithm typically produces a metric first in a null coordinate system. One then attempts to pass to a Boyer--Lindquist-type chart in which the \(tr\) and \(r\phi\) components vanish. For the off-shell metric generated in Sec. \ref{sec2}, we consider an Eddington--Finkelstein-type coordinate system \((u,r,\theta,\phi)\) and seek a transformation of the form
\begin{equation}
du
=
dt
+
\lambda(r,\theta)\,dr,
\quad
d\phi
=
d\varphi
+
\chi(r,\theta)\,dr .
\label{4.11}
\end{equation}
This is not yet a legitimate coordinate transformation. For \(t\) and \(\varphi\) to exist as coordinates obtained by integrating Eq. \eqref{4.11}, the one-forms on the right-hand side must be exact. Hence
\begin{equation}
\partial_\theta\lambda
=
0,
\quad
\partial_\theta\chi
=
0 .
\label{4.12}
\end{equation}
Only in this case may Eq. \eqref{4.11} be written as
\begin{equation}
du
=
dt+\lambda(r)\,dr,
\quad
d\phi
=
d\varphi+\chi(r)\,dr .
\label{4.13}
\end{equation}
This distinction is important in Newman--Janis constructions: one can always algebraically choose \(\lambda(r,\theta)\) and \(\chi(r,\theta)\) to cancel selected metric components, but such a choice is not a genuine Boyer--Lindquist transformation unless Eq. \eqref{4.12} holds. This issue is emphasized in modern discussions of Newman--Janis algorithms and non-complexification variants \cite{Erbin:2014aya,Azreg-Ainou:2014pra}.

Let the relevant part of the off-shell metric in \((u,r,\theta,\phi)\) coordinates be
\begin{align}
ds^2
&=
g_{uu}\,du^2
+
2g_{ur}\,du\,dr
+
2g_{u\phi}\,du\,d\phi
+
2g_{r\phi}\,dr\,d\phi
\nn&+
g_{\phi\phi}\,d\phi^2
+
g_{rr}\,dr^2
+
g_{\theta\theta}\,d\theta^2 .
\label{4.14}
\end{align}
All components in Eq. \eqref{4.14} depend on \(r\), \(\theta\), and the rotation parameter \(a\), but not on \(u\) or \(\phi\). Under Eq. \eqref{4.11}, the mixed radial components in the new coordinates are
\begin{equation}
g_{tr}^{\rm new}
=
g_{uu}\lambda
+
g_{ur}
+
g_{u\phi}\chi ,
\label{4.15}
\end{equation}
and
\begin{equation}
g_{r\varphi}^{\rm new}
=
g_{u\phi}\lambda
+
g_{r\phi}
+
g_{\phi\phi}\chi .
\label{4.16}
\end{equation}
The Boyer--Lindquist algebraic conditions are therefore
\begin{equation}
g_{tr}^{\rm new}
=
0,
\quad
g_{r\varphi}^{\rm new}
=
0 .
\label{4.17}
\end{equation}
When the determinant
\begin{equation}
\mathscr{D}
=
g_{uu}g_{\phi\phi}
-
g_{u\phi}^{\,2}
\label{4.18}
\end{equation}
is nonzero, Eqs. \eqref{4.15}--\eqref{4.17} yield
\begin{equation}
\lambda(r,\theta)
=
\frac{
-g_{ur}g_{\phi\phi}
+
g_{u\phi}g_{r\phi}
}
{\mathscr{D}},
\label{4.19}
\end{equation}
and
\begin{equation}
\chi(r,\theta)
=
\frac{
g_{u\phi}g_{ur}
-
g_{uu}g_{r\phi}
}
{\mathscr{D}} .
\label{4.20}
\end{equation}
Equations \eqref{4.19} and \eqref{4.20} are algebraic consequences of the coordinate ansatz \eqref{4.11}. They provide the candidate radial shifts needed to remove \(g_{tr}\) and \(g_{r\varphi}\).

The Boyer--Lindquist coordinate-integrability conditions are then
\begin{align}
\partial_\theta
\left[
\frac{
-g_{ur}g_{\phi\phi}
+
g_{u\phi}g_{r\phi}
}
{g_{uu}g_{\phi\phi}-g_{u\phi}^{\,2}}
\right]
=
0,
\nn
\partial_\theta
\left[
\frac{
g_{u\phi}g_{ur}
-
g_{uu}g_{r\phi}
}
{g_{uu}g_{\phi\phi}-g_{u\phi}^{\,2}}
\right]
=
0 .
\label{4.21}
\end{align}
Equation \eqref{4.21} is the local test for the existence of a Boyer--Lindquist-type transformation of the form \eqref{4.13}, away from the locus \(\mathscr{D}=0\). The surface \(\mathscr{D}=0\) is the degeneracy locus of the Killing-sector metric. In a Kerr-like spacetime this includes physically significant hypersurfaces associated with stationary-limit behavior, so Eq. \eqref{4.21} should be interpreted as a local condition on patches where the transformation is nonsingular.

If Eq. \eqref{4.21} holds, the metric can be written locally in the Boyer--Lindquist-type form
\begin{align}
ds^2
&=
g_{tt}^{\rm BL}(r,\theta)\,dt^2
+
2g_{t\varphi}^{\rm BL}(r,\theta)\,dt\,d\varphi
+
g_{\varphi\varphi}^{\rm BL}(r,\theta)\,d\varphi^2
\nn&+
g_{rr}^{\rm BL}(r,\theta)\,dr^2
+
g_{\theta\theta}^{\rm BL}(r,\theta)\,d\theta^2 .
\label{4.22}
\end{align}
The original Boyer--Lindquist coordinates were introduced to display the Kerr metric in a form adapted to stationarity and axial symmetry \cite{Boyer:1966qh}. In the present paper, however, Eq. \eqref{4.22} is not assumed a priori. It is a geometrical property that an off-shell Newman--Janis deformation may or may not possess. It is worth noting that, in the non-complexification procedure of Ref.~\cite{Azreg-Ainou:2014pra}, the functions \(\lambda(r)\) and \(\chi(r)\) are expressed solely in terms of the static seed metric components, so that the Boyer--Lindquist form \eqref{4.22} is always achievable within the assumptions of that prescription.

The distinction between circularity and Boyer--Lindquist integrability should also be kept explicit. Circularity is the invariant Frobenius condition in Eq. \eqref{4.7}. Boyer--Lindquist integrability is a coordinate-level condition requiring that the particular radial shifts \(\lambda\) and \(\chi\) be functions of \(r\) alone. A circular spacetime may require a different meridional coordinate choice before taking the simple form \eqref{4.22}, while a non-circular spacetime cannot be made circular by a coordinate transformation that preserves the Killing directions. Thus, Boyer--Lindquist failure is a useful consistency check, but the covariant circularity obstruction is the deeper geometrical obstruction.

\subsection{Axis regularity, elementary flatness, and equatorial reflection}

Circularity and Boyer--Lindquist integrability are not the only geometrical restrictions on a physically admissible rotating candidate. A stationary and axisymmetric spacetime must also have a regular symmetry axis, must be free of conical defects unless such defects are part of the intended physical model, and must obey the prescribed discrete reflection symmetry when the rotating object is assumed to be equatorially symmetric.

Let
\begin{equation}
\eta^\mu
=
\left(\frac{\partial}{\partial\phi}\right)^\mu
\label{eq:axis_killing_vector}
\end{equation}
be the axial Killing vector. The rotation axis is the fixed-point set of the axial action,
\begin{equation}
\eta^2
\equiv
g_{\mu\nu}\eta^\mu\eta^\nu
=
g_{\phi\phi}
=
0 .
\label{eq:axis_fixed_point_set}
\end{equation}
Regularity on this set requires that the metric be smooth when written in local Cartesian coordinates transverse to the axis. In coordinates adapted to the usual polar angle, this implies
\begin{equation}
g_{\phi A}
=
O(\sin^2\theta),
\quad
A\neq\phi,
\quad
\theta\to0,\pi ,
\label{eq:axis_cross_term_falloff}
\end{equation}
and
\begin{equation}
g_{\phi\phi}
=
O(\sin^2\theta).
\label{eq:gphiphi_axis_falloff}
\end{equation}
These conditions ensure that the axial coordinate degenerates smoothly rather than producing a string-like singularity in the metric components.

The absence of a conical singularity is the stronger elementary-flatness condition. In invariant form it is
\begin{equation}
\lim_{\eta^2\to0}
\frac{
\nabla_\mu(\eta^2)\nabla^\mu(\eta^2)
}
{4\eta^2}
=
1 .
\label{eq:elementary_flatness_sec4}
\end{equation}
Equivalently, in the angular coordinate used in the Newman--Janis ansatz,
\begin{equation}
\lim_{\theta\to0,\pi}
\frac{
g_{\phi\phi}
}
{
g_{\theta\theta}\sin^2\theta
}
=
1 .
\label{eq:no_conical_coordinate_sec4}
\end{equation}
It is useful to define the corresponding axis-defect functions
\begin{equation}
\mathcal{A}_{\pm}(r;a)
\equiv
\lim_{\theta\to\theta_\pm}
\left[
\frac{
g_{\phi\phi}
}
{
g_{\theta\theta}\sin^2\theta
}
-
1
\right],
\quad
\theta_+=0,
\quad
\theta_-=\pi .
\label{eq:axis_defect_functions}
\end{equation}
A nonzero \(\mathcal{A}_{\pm}\) indicates a conical excess or deficit on the corresponding part of the axis. The associated deficit/excess angle may be written as
\begin{equation}
\delta_\pm(r;a)
=
2\pi
\left[
1-
\lim_{\theta\to\theta_\pm}
\sqrt{
\frac{
g_{\phi\phi}
}
{
g_{\theta\theta}\sin^2\theta
}
}
\right].
\label{eq:conical_defect_angle}
\end{equation}
Thus the no-conical-singularity requirement is
\begin{equation}
\mathcal{A}_{+}=0,
\quad
\mathcal{A}_{-}=0 .
\label{eq:no_conical_requirement}
\end{equation}

For the ansatz \eqref{2.15}, Eq. \eqref{eq:axis_defect_functions} reduces to
\begin{equation}
\mathcal{A}_{\pm}(r;a)
=
\lim_{\theta\to\theta_\pm}
\left[
\frac{G(r,\theta;a)}{E(r,\theta;a)}
-
1
\right].
\label{eq:axis_defect_EG}
\end{equation}
Therefore, elementary flatness is equivalent to
\begin{equation}
G(r,0;a)=E(r,0;a),
\quad
G(r,\pi;a)=E(r,\pi;a).
\label{eq:EG_axis_equal_sec4}
\end{equation}
This condition is automatically satisfied by the Kerr metric in Boyer--Lindquist form, but it is not guaranteed by an arbitrary Newman--Janis-type deformation.

Finally, for an equatorially symmetric rotating candidate, reflection through the equatorial plane is imposed as the discrete isometry
\begin{equation}
\mathcal{R}_{\rm eq}:\theta\mapsto\pi-\theta .
\label{eq:req_map_sec4}
\end{equation}
For the ansatz \eqref{2.15}, the reflection residuals are
\begin{equation}
\mathcal{R}_{\rm eq}[X](r,\theta;a)
\equiv
X(r,\pi-\theta;a)-X(r,\theta;a),
\label{eq:req_residual_functions}
\end{equation}
where $X\in\{A,B,C,D,E,G\}.$ Equatorial reflection symmetry requires
\begin{equation}
\mathcal{R}_{\rm eq}[X]=0
\quad
\hbox{for all}
\quad
X\in\{A,B,C,D,E,G\}.
\label{eq:req_zero_sec4}
\end{equation}
Equivalently, these coefficient functions must be even functions of \(\cos\theta\). In a more general ansatz containing components with a single \(\theta\)-index, those components must be odd under \(\theta\mapsto\pi-\theta\), while components with no \(\theta\)-index or two \(\theta\)-indices must be even. The equatorial reflection condition therefore fixes the allowed parity of both the Newman--Janis output and any residual completion.

\subsection{Geometrical obstruction functions}
We now convert the circularity and Boyer--Lindquist conditions into explicit obstruction functions. Let \(\epsilon_{\mu\nu\rho\sigma}\) be the volume form associated with \(g_{\mu\nu}\). In four dimensions, the two circularity four-forms in Eq. \eqref{4.6} are equivalent to the scalar densities
\begin{equation}
\mathcal{C}_{\xi}
=
\epsilon^{\mu\nu\rho\sigma}
\xi_\mu
\eta_\nu
\nabla_\rho \xi_\sigma,
\label{4.23}
\end{equation}
and
\begin{equation}
\mathcal{C}_{\eta}
=
\epsilon^{\mu\nu\rho\sigma}
\xi_\mu
\eta_\nu
\nabla_\rho \eta_\sigma .
\label{4.24}
\end{equation}
The spacetime is circular on a region where the Killing fields are linearly independent if and only if
\begin{equation}
\mathcal{C}_{\xi}
=
0,
\quad
\mathcal{C}_{\eta}
=
0 .
\label{4.25}
\end{equation}
Equations \eqref{4.23}--\eqref{4.25} are simply the Hodge-dual form of the Frobenius circularity conditions in Eq. \eqref{4.7}.

For a metric written in the Newman--Janis coordinates of Eq. \eqref{4.14}, we define the Boyer--Lindquist obstruction functions
\begin{equation}
\mathcal{B}_{\lambda}
=
\partial_\theta\lambda,
\quad
\mathcal{B}_{\chi}
=
\partial_\theta\chi ,
\label{4.26}
\end{equation}
where \(\lambda\) and \(\chi\) are the algebraic candidates in Eqs. \eqref{4.19} and \eqref{4.20}. Explicitly,
\begin{equation}
\mathcal{B}_{\lambda}
=
\partial_\theta
\left[
\frac{
-g_{ur}g_{\phi\phi}
+
g_{u\phi}g_{r\phi}
}
{g_{uu}g_{\phi\phi}-g_{u\phi}^{\,2}}
\right],
\label{4.27}
\end{equation}
and
\begin{equation}
\mathcal{B}_{\chi}
=
\partial_\theta
\left[
\frac{
g_{u\phi}g_{ur}
-
g_{uu}g_{r\phi}
}
{g_{uu}g_{\phi\phi}-g_{u\phi}^{\,2}}
\right].
\label{4.28}
\end{equation}
The vanishing of \(\mathcal{B}_{\lambda}\) and \(\mathcal{B}_{\chi}\) is equivalent to the existence of a Boyer--Lindquist-type transformation of the restricted form \eqref{4.13}.

The geometrical obstruction tuple is then
\begin{equation}
\mathfrak{G}
\left[g^{\rm NJA}\right]
=
\left(
\mathcal{C}_{\xi},
\mathcal{C}_{\eta},
\mathcal{B}_{\lambda},
\mathcal{B}_{\chi},
\mathcal{A}_{+},
\mathcal{A}_{-},
\mathcal{R}_{\rm eq}
\right).
\label{4.29}
\end{equation}
We say that the off-shell Newman--Janis deformation is geometrically admissible, in the circular, Boyer--Lindquist-integrable, axis-regular, elementary-flat, and equatorially symmetric sense, when
\begin{equation}
\boxed{
\mathfrak{G}
\left[g^{\rm NJA}\right]
=
0 .
}
\label{4.30}
\end{equation}
Equivalently,
\begin{align}
\mathcal{C}_{\xi}
=
0,
\quad
\mathcal{C}_{\eta}
=
0,
\quad
\mathcal{B}_{\lambda}
=
0,
\quad
\mathcal{B}_{\chi}
=
0,
\nn
\mathcal{A}_{+}
=
0,
\quad
\mathcal{A}_{-}
=
0,
\quad
\mathcal{R}_{\rm eq}
=
0 .
\label{4.31}
\end{align}
This is the strengthened geometrical admissibility condition used in the rest of the paper. It separates local circularity and coordinate integrability from the global regularity requirements associated with the axial fixed-point set and the imposed equatorial reflection symmetry.

Because the Newman--Janis output reduces to the static seed at \(a=0\), the obstruction functions admit a rotation expansion,
\begin{equation}
\mathcal{C}_{\xi}
=
a\,\mathcal{C}_{\xi}^{(1)}
+
a^2\mathcal{C}_{\xi}^{(2)}
+
a^3\mathcal{C}_{\xi}^{(3)}
+
O(a^4),
\label{4.32}
\end{equation}
\begin{equation}
\mathcal{C}_{\eta}
=
a\,\mathcal{C}_{\eta}^{(1)}
+
a^2\mathcal{C}_{\eta}^{(2)}
+
a^3\mathcal{C}_{\eta}^{(3)}
+
O(a^4),
\label{4.33}
\end{equation}
and
\begin{align}
\mathcal{B}_{\lambda}
=
a\,\mathcal{B}_{\lambda}^{(1)}
+
a^2\mathcal{B}_{\lambda}^{(2)}
+
a^3\mathcal{B}_{\lambda}^{(3)}
+
O(a^4),
\nn
\mathcal{B}_{\chi}
=
a\,\mathcal{B}_{\chi}^{(1)}
+
a^2\mathcal{B}_{\chi}^{(2)}
+
a^3\mathcal{B}_{\chi}^{(3)}
+
O(a^4).
\label{4.34}
\end{align}
The leading geometrical obstruction order is
\begin{equation}
n_{\rm geom}
=
\min
\left\{
n:
\mathcal{C}_{\xi}^{(n)}
\neq0
\ \hbox{or}\
\mathcal{C}_{\eta}^{(n)}
\neq0
\ \hbox{or}\
\mathcal{B}_{\lambda}^{(n)}
\neq0
\ \hbox{or}\
\mathcal{B}_{\chi}^{(n)}
\neq0
\right\}.
\label{4.35}
\end{equation}
If \(n_{\rm geom}\) is finite, a residual completion must begin no later than order \(a^{n_{\rm geom}}\) in the corresponding metric sector. If no such correction is allowed within the chosen ansatz class, the Newman--Janis output fails geometrically even before the dynamical field equations are considered.

For practical computations, one may combine the scalar obstructions into a single nonnegative indicator,
\begin{equation}
\mathscr{G}
=
\mathcal{C}_{\xi}^{\,2}
+
\mathcal{C}_{\eta}^{\,2}
+
\mathcal{B}_{\lambda}^{\,2}
+
\mathcal{B}_{\chi}^{\,2}.
\label{4.36}
\end{equation}
The scalar \(\mathscr{G}\) is not itself a covariant invariant because \(\mathcal{B}_{\lambda}\) and \(\mathcal{B}_{\chi}\) depend on the chosen Newman--Janis coordinate gauge. Its zero set, however, encodes the simultaneous satisfaction of the circularity and Boyer--Lindquist integrability conditions in that gauge. The covariant part of the geometrical obstruction is carried by \(\mathcal{C}_{\xi}\) and \(\mathcal{C}_{\eta}\); the coordinate-integrability part is carried by \(\mathcal{B}_{\lambda}\) and \(\mathcal{B}_{\chi}\).

This distinction will be important in Sec. \ref{sec5}. If \(\mathcal{C}_{\xi}\) or \(\mathcal{C}_{\eta}\) is nonzero, the failure is invariant and cannot be repaired by a Boyer--Lindquist coordinate redefinition alone. If \(\mathcal{C}_{\xi}=\mathcal{C}_{\eta}=0\) but either \(\mathcal{B}_{\lambda}\) or \(\mathcal{B}_{\chi}\) is nonzero, the spacetime may still be circular, but the Newman--Janis coordinates do not admit the simple radial Boyer--Lindquist transformation assumed in Eq. \eqref{4.13}. The residual-completion problem must therefore distinguish invariant geometrical obstructions from coordinate-gauge obstructions.

\section{Residual Completion of the Newman--Janis Ansatz} \label{sec5}
\subsection{Corrected metric and matter fields}
The obstruction tensor introduced in Sec. \ref{sec3} converts the Newman--Janis failure problem into a residual problem. We now ask whether the residual can be cancelled by correcting the off-shell Newman--Janis metric and its associated trial matter fields within a controlled ansatz class. Let
\begin{equation}
\left(
g_{\mu\nu}^{\rm NJA}(a),
\Psi^{\rm trial}(a)
\right)
\label{5.1}
\end{equation}
be the Newman--Janis-type image of the static seed. We introduce corrected fields by writing
\begin{equation}
g_{\mu\nu}(a)
=
g_{\mu\nu}^{\rm NJA}(a)
+
\delta g_{\mu\nu}(a),
\quad
\Psi(a)
=
\Psi^{\rm trial}(a)
+
\delta\Psi(a).
\label{5.2}
\end{equation}
The corrections are required to vanish in the nonrotating limit,
\begin{equation}
\lim_{a\to0}
\delta g_{\mu\nu}(a)
=
0,
\quad
\lim_{a\to0}
\delta\Psi(a)
=
0.
\label{5.3}
\end{equation}
Thus, residual completion does not change the static seed. It only repairs those rotating sectors that are not fixed consistently by the Newman--Janis prescription.

The metric correction is expanded in powers of the rotation parameter,
\begin{equation}
\delta g_{\mu\nu}(a)
=
a\,q_{\mu\nu}^{(1)}
+
a^2 q_{\mu\nu}^{(2)}
+
a^3 q_{\mu\nu}^{(3)}
+
O(a^4),
\label{5.4}
\end{equation}
and the matter correction is expanded similarly,
\begin{equation}
\delta\Psi(a)
=
a\,\psi^{(1)}
+
a^2\psi^{(2)}
+
a^3\psi^{(3)}
+
O(a^4).
\label{5.5}
\end{equation}
The tensors \(q_{\mu\nu}^{(n)}\) and fields \(\psi^{(n)}\) are not arbitrary. They must preserve stationarity, axisymmetry, equatorial reflection symmetry, and the chosen asymptotic or near-horizon boundary conditions. Under reflection across the equatorial plane and reversal of the sense of rotation, the axial sector is odd in \(a\), while the polar sector is even in \(a\). This is the same parity organization used in the slow-rotation expansion of relativistic compact objects, where first order in rotation describes frame dragging and second order describes monopolar and quadrupolar deformation \cite{Hartle:1967he,Hartle:1968si}.

For a circular stationary and axisymmetric completion, the correction may be decomposed into axial and polar sectors,
\begin{equation}
q_{\mu\nu}^{(n)}
=
q_{\mu\nu}^{(n),{\rm ax}}
+
q_{\mu\nu}^{(n),{\rm pol}} .
\label{5.6}
\end{equation}
The axial sector contains the rotational cross terms, such as \(q_{t\phi}^{(1)}\), while the polar sector contains corrections to \(g_{tt}\), \(g_{rr}\), \(g_{\theta\theta}\), and \(g_{\phi\phi}\). In a slow-rotation treatment around a spherical seed, this decomposition is the stationary version of the usual odd-parity and even-parity perturbation split used in black-hole perturbation theory \cite{Regge:1957td,Zerilli:1970se,Martel:2005ir}.

A useful representation of the correction is through an angular harmonic expansion. For reflection-symmetric configurations, the even-parity sector at second order may be written as
\begin{equation}
q_{\mu\nu}^{(2),{\rm pol}}(r,\theta)
=
q_{\mu\nu}^{(2,0)}(r)P_0(\cos\theta)
+
q_{\mu\nu}^{(2,2)}(r)P_2(\cos\theta),
\label{5.7}
\end{equation}
where \(P_\ell\) are Legendre polynomials. The \(\ell=0\) sector renormalizes the monopolar part of the geometry, while the \(\ell=2\) sector encodes the quadrupolar deformation. At first order, the stationary axial sector is represented by
\begin{equation}
q_{t\phi}^{(1)}(r,\theta)
=
-\omega_1(r)R^2(r)\sin^2\theta ,
\label{5.8}
\end{equation}
where \(\omega_1(r)\) is the leading frame-dragging function. Equation \eqref{5.8} is the standard slow-rotation form of the axial perturbation around a spherical background \cite{Hartle:1967he}. In the present paper it is not imposed universally; it is used only when the leading obstruction lies in the dipolar frame-dragging channel.

Matter corrections must be chosen consistently with the matter model. For an electromagnetic seed, one may write
\begin{align}
A_\mu(a)
=
A_\mu^{\rm trial}(a)
+
\delta A_\mu(a),
\nn
\delta A_\mu(a)
=
a\,\alpha_\mu^{(1)}
+
a^2\alpha_\mu^{(2)}
+
O(a^3).
\label{5.9}
\end{align}
For a scalar seed,
\begin{equation}
\Phi(a)
=
\Phi^{\rm trial}(a)
+
\delta\Phi(a),
\quad
\delta\Phi(a)
=
a\,\varphi^{(1)}
+
a^2\varphi^{(2)}
+
O(a^3).
\label{5.10}
\end{equation}
For an effective anisotropic fluid, the correction acts not only on the eigenvalues of \(T^\mu{}_\nu\) but also on the orthonormal tetrad used to define the density and principal pressures. This distinction is essential because the same metric can be made to solve Einstein's equations by defining \(T_{\mu\nu}^{\rm eff}=G_{\mu\nu}/8\pi\), but that effective source need not preserve the physical matter interpretation of the static seed.

The corrected fields are required to satisfy
\begin{align}
\mathfrak{E}_{\mu\nu}
\left[
g^{\rm NJA}+\delta g,
\Psi^{\rm trial}+\delta\Psi
\right]
=
0,
\nn
\mathfrak{S}_I
\left[
g^{\rm NJA}+\delta g,
\Psi^{\rm trial}+\delta\Psi
\right]
=
0 .
\label{5.11}
\end{align}
Equation \eqref{5.11} is the exact residual-completion condition. It is generally nonlinear. The purpose of the perturbative expansion is to convert it into a sequence of linear solvability problems ordered by powers of \(a\).

\subsection{Linearized residual-completion equation}
Let the Newman--Janis obstruction be expanded as
\begin{equation}
\mathcal{O}_{\mu\nu}^{\rm NJA}(a)
=
a^{n_\star}
\mathcal{O}_{\mu\nu}^{(n_\star)}
+
O(a^{n_\star+1}),
\label{5.12}
\end{equation}
where \(n_\star\) is the leading obstruction order defined in Sec. \ref{sec3}. We now expand the corrected metric equation around the off-shell Newman--Janis configuration. To first order in the correction, one obtains
\begin{align}
&\mathfrak{E}_{\mu\nu}
\left[
g^{\rm NJA}+\delta g,
\Psi^{\rm trial}+\delta\Psi
\right]
=
\mathfrak{E}_{\mu\nu}
\left[
g^{\rm NJA},
\Psi^{\rm trial}
\right]
\nn&+
\mathcal{L}_{\mu\nu}^{\ \ \alpha\beta}
\left[
g^{\rm NJA},\Psi^{\rm trial}
\right]
\delta g_{\alpha\beta}
+
\mathcal{J}_{\mu\nu}
\left[
g^{\rm NJA},\Psi^{\rm trial}
\right]\delta\Psi
+
Q_{\mu\nu},
\label{5.13}
\end{align}
where \(Q_{\mu\nu}\) is at least quadratic in \((\delta g,\delta\Psi)\). The operator \(\mathcal{L}_{\mu\nu}^{\ \ \alpha\beta}\) is the metric linearization of the field equations, and \(\mathcal{J}_{\mu\nu}\) is the part produced by the matter-field correction. Linearization of the Einstein tensor and the corresponding gauge dependence of metric perturbations are standard features of gravitational perturbation theory \cite{Wald:1984rg,Regge:1957td,Martel:2005ir}.

To obtain the leading completion equation, we now insert
\begin{equation}
\mathcal{O}_{\mu\nu}^{\rm NJA}(a)
=
a^{n_\star}\mathcal{O}_{\mu\nu}^{(n_\star)}
+
O(a^{n_\star+1}),
\end{equation}
together with
\begin{equation}
\delta g_{\mu\nu}
=
a^{n_\star}q_{\mu\nu}^{(n_\star)}
+
O(a^{n_\star+1}),
\quad
\delta\Psi
=
a^{n_\star}\psi^{(n_\star)}
+
O(a^{n_\star+1}).
\end{equation}
The linearized metric and matter terms in Eq. \eqref{5.13} therefore contribute at order \(a^{n_\star}\), whereas \(Q_{\mu\nu}\), being at least quadratic in \((\delta g,\delta\Psi)\), starts at order \(a^{2n_\star}\). Hence \(Q_{\mu\nu}\) does not contribute to the coefficient of \(a^{n_\star}\).

At the leading obstruction order, the quadratic terms in Eq. \eqref{5.13} do not contribute. Since
\begin{equation}
\mathfrak{E}_{\mu\nu}
\left[
g^{\rm NJA},
\Psi^{\rm trial}
\right]
=
\mathcal{O}_{\mu\nu}^{\rm NJA},
\label{5.14}
\end{equation}
the order \(a^{n_\star}\) part of Eq. \eqref{5.11} gives
\begin{equation}
\mathcal{O}_{\mu\nu}^{(n_\star)}
+
\mathcal{L}_{\mu\nu}^{(0)}
\left[
q^{(n_\star)}
\right]
+
\mathcal{J}_{\mu\nu}^{(0)}
\left[
\psi^{(n_\star)}
\right]
=
0 .
\label{5.15}
\end{equation}
Here \(\mathcal{L}^{(0)}\) and \(\mathcal{J}^{(0)}\) are evaluated on the static seed when the leading obstruction is computed relative to the seed background. More generally, if lower-order Newman--Janis sectors are retained exactly, the same equation is evaluated on the lower-order off-shell background. We define the combined residual-completion operator by
\begin{equation}
\mathbb{L}_{\rm NJA}^{(n_\star)}
\left[
q^{(n_\star)},\psi^{(n_\star)}
\right]
\equiv
\mathcal{L}_{\mu\nu}^{(0)}
\left[
q^{(n_\star)}
\right]
+
\mathcal{J}_{\mu\nu}^{(0)}
\left[
\psi^{(n_\star)}
\right].
\label{5.16}
\end{equation}
With the obstruction convention of Sec. \ref{sec3}, the leading residual-completion equation is therefore
\begin{equation}
\boxed{
\mathbb{L}_{\rm NJA}^{(n_\star)}
\left[
q^{(n_\star)},\psi^{(n_\star)}
\right]
=
-\mathcal{O}_{\mu\nu}^{(n_\star)} .
}
\label{5.17}
\end{equation}
This is the central equation of the residual-completion framework. It says that the Newman--Janis obstruction must lie in the image of the linearized completion operator, up to the sign fixed by the definition of \(\mathcal{O}_{\mu\nu}^{\rm NJA}\).

For Einstein gravity with a fixed matter correction, the metric part of the operator is the linearized Einstein operator. If
\begin{equation}
h_{\mu\nu}
=
q_{\mu\nu}^{(n_\star)},
\quad
h
=
g^{(0)\mu\nu}h_{\mu\nu},
\label{5.18}
\end{equation}
then the first variation of the Einstein tensor may be written as
\begin{equation}
\delta G_{\mu\nu}[h]
=
\delta R_{\mu\nu}[h]
-
\frac{1}{2}g_{\mu\nu}^{(0)}\delta R[h]
-
\frac{1}{2}h_{\mu\nu}R^{(0)}
+
\frac{1}{2}g_{\mu\nu}^{(0)}h^{\alpha\beta}R_{\alpha\beta}^{(0)} .
\label{5.19}
\end{equation}
The corresponding Ricci variation is
\begin{equation}
\delta R_{\mu\nu}[h]
=
\frac{1}{2}
\left(
-\nabla^2 h_{\mu\nu}
-\nabla_\mu\nabla_\nu h
+\nabla_\alpha\nabla_\mu h^\alpha{}_\nu
+\nabla_\alpha\nabla_\nu h^\alpha{}_\mu
\right),
\label{5.20}
\end{equation}
where all covariant derivatives are taken with respect to the seed metric. Equations \eqref{5.19} and \eqref{5.20} are the standard linearized curvature formulas \cite{Wald:1984rg,Carroll_2019}. In Einstein gravity, Eq. \eqref{5.17} becomes
\begin{equation}
\delta G_{\mu\nu}
\left[
q^{(n_\star)}
\right]
-
8\pi
\delta T_{\mu\nu}
\left[
q^{(n_\star)},\psi^{(n_\star)}
\right]
=
-\mathcal{O}_{\mu\nu}^{(n_\star)} .
\label{5.21}
\end{equation}

The completion equation is meaningful only when its right-hand side satisfies the compatibility identities inherited from diffeomorphism invariance. On the seed background, the contracted Bianchi identity implies
\begin{equation}
\nabla_{(0)}^\mu
\delta G_{\mu\nu}[h]
=
0
\label{5.22}
\end{equation}
whenever the linearized equations are evaluated consistently around a solution. More generally, for a diffeomorphism-invariant Lagrangian theory, the corresponding Noether identity relates the divergence of the linearized metric equations to the linearized matter equations \cite{Iyer:1994ys}. Thus, a necessary compatibility condition for Eq. \eqref{5.17} is
\begin{equation}
\nabla_{(0)}^\mu
\mathcal{O}_{\mu\nu}^{(n_\star)}
=
0
\label{5.23}
\end{equation}
when the matter obstruction at the same order is also cancelled. If Eq. \eqref{5.23} fails, then no metric correction alone can cancel the obstruction. One must either modify the matter correction, enlarge the ansatz, or reject the Newman--Janis deformation within the chosen theory.

In operator language, let \(\mathbb{L}_{\rm NJA}^{(n_\star)\dagger}\) denote the formal adjoint of the completion operator with respect to the chosen inner product and boundary conditions. A necessary Fredholm-type solvability condition for Eq. \eqref{5.17} is
\begin{equation}
\left\langle
Z,
\mathcal{O}^{(n_\star)}
\right\rangle
=
0
\quad
\hbox{for every}
\quad
Z\in
{\rm Ker}
\left(
\mathbb{L}_{\rm NJA}^{(n_\star)\dagger}
\right).
\label{5.24}
\end{equation}
This is the standard adjoint-kernel compatibility condition for an inhomogeneous linear problem \cite{Kato_1995,1972}. In the present setting, it supplies a precise obstruction criterion: if the leading Newman--Janis residual has a nonzero projection onto the cokernel of the completion operator, then no residual completion exists within the selected function space and boundary conditions.

When Eq. \eqref{5.17} is solvable, the solution is not unique unless gauge, boundary, and minimality conditions are imposed. If \(X\) belongs to the kernel of \(\mathbb{L}_{\rm NJA}^{(n_\star)}\), then
\begin{equation}
\left(
q^{(n_\star)},\psi^{(n_\star)}
\right)
\longmapsto
\left(
q^{(n_\star)},\psi^{(n_\star)}
\right)
+
X
\label{5.25}
\end{equation}
generates another solution of the same inhomogeneous equation. Some kernel elements represent physical homogeneous perturbations, such as shifts of mass, charge, or angular momentum. Others represent gauge freedom. The next subsection fixes the latter and restricts the former.

\subsection{Minimality and gauge fixing}
Metric perturbations are defined modulo infinitesimal diffeomorphisms. Under
\begin{equation}
x^\mu
\longmapsto
x^\mu
-
\varepsilon\,\zeta^\mu ,
\label{5.26}
\end{equation}
a metric perturbation transforms as
\begin{equation}
q_{\mu\nu}^{(n)}
\longmapsto
q_{\mu\nu}^{(n)}
+
\nabla_\mu\zeta_\nu
+
\nabla_\nu\zeta_\mu .
\label{5.27}
\end{equation}
Equation \eqref{5.27} is the standard gauge transformation law for metric perturbations \cite{Wald:1984rg,Regge:1957td}. Therefore, a residual completion is not a unique tensor until a gauge condition has been imposed.

A covariant possibility is the de Donder condition on the trace-reversed perturbation,
\begin{equation}
\bar q_{\mu\nu}^{(n)}
=
q_{\mu\nu}^{(n)}
-
\frac{1}{2}g_{\mu\nu}^{(0)}q^{(n)},
\quad
q^{(n)}
=
g_{(0)}^{\alpha\beta}q_{\alpha\beta}^{(n)},
\label{5.28}
\end{equation}
with
\begin{equation}
\mathcal{G}_\nu[q^{(n)}]
\equiv
\nabla^\mu
\bar q_{\mu\nu}^{(n)}
=
0 .
\label{5.29}
\end{equation}
The de Donder condition is widely used because it renders the principal part of the linearized Einstein operator hyperbolic in Lorentzian problems and elliptic after stationary reduction under suitable assumptions \cite{Wald:1984rg,Carroll_2019}. For a spherical seed and a multipolar decomposition, one may instead use a Regge--Wheeler-type gauge adapted to axial and polar perturbations \cite{Regge:1957td,Zerilli:1970se,Martel:2005ir}. The choice of gauge is not part of the physics; it is part of the definition of the completion problem.

We impose a general gauge condition
\begin{equation}
\mathcal{G}_\mu
\left[
q^{(n_\star)}
\right]
=
0 .
\label{5.30}
\end{equation}
The gauge-fixed residual-completion problem is then
\begin{equation}
\left\{
\begin{array}{rcl}
\mathbb{L}_{\rm NJA}^{(n_\star)}
\left[
q^{(n_\star)},\psi^{(n_\star)}
\right]
&=&
-\mathcal{O}_{\mu\nu}^{(n_\star)},
\\[4pt]
\mathcal{G}_\mu
\left[
q^{(n_\star)}
\right]
&=&
0 .
\end{array}
\right.
\label{5.31}
\end{equation}
Boundary conditions complete the problem. For asymptotically flat black-hole applications, the correction should not spoil the asymptotic charges unless a charge renormalization is intended. Thus one may impose
\begin{equation}
q_{\mu\nu}^{(n_\star)}
=
O(r^{-1})
\quad
\hbox{or faster as}
\quad
r\to\infty ,
\label{5.32}
\end{equation}
with the precise falloff determined by the multipole sector under consideration. Near a regular horizon, the correction should remain finite in a horizon-penetrating coordinate system,
\begin{equation}
\left|
q_{\mu\nu}^{(n_\star)}
\right|
<
\infty
\quad
\hbox{as}
\quad
r\to r_h ,
\label{5.33}
\end{equation}
where \(r_h\) is the seed or corrected horizon location to the order considered. For regular seeds, including regular black holes and wormholes, one also imposes regularity at the center or at the corresponding regular core or throat,
\begin{equation}
\left|
q_{\mu\nu}^{(n_\star)}
\right|
<
\infty
\quad
\hbox{as}
\quad
r\to0 ,
\label{5.34}
\end{equation}
in an appropriate regular frame.

The boundary operator \(\mathfrak{B}\) must also include the axial and reflection conditions introduced in Sec. \ref{sec4}. At the symmetry axis, a completion must preserve the falloff
\begin{equation}
q_{\phi A}^{(n_\star)}
=
O(\sin^2\theta),
\quad
A\neq\phi,
\quad
\theta\to0,\pi ,
\label{eq:completion_axis_cross_terms}
\end{equation}
and must not introduce a conical defect. If the background already satisfies elementary flatness, then the order-\(a^{n_\star}\) correction must satisfy
\begin{equation}
\left.
\left[
\frac{
q_{\phi\phi}^{(n_\star)}
}
{
\sin^2\theta
}
-
q_{\theta\theta}^{(n_\star)}
\right]
\right|_{\theta=0,\pi}
=
0 .
\label{eq:completion_no_conical_condition}
\end{equation}
For the parametrization \(g_{\theta\theta}=E\) and \(g_{\phi\phi}=\sin^2\theta\,G\), this is the perturbative form of
\begin{equation}
G(r,0;a)=E(r,0;a),
\quad
G(r,\pi;a)=E(r,\pi;a).
\label{eq:completion_EG_axis}
\end{equation}
Equatorial reflection symmetry further restricts the correction functions to the same parity class as the uncorrected ansatz:
\begin{equation}
q_{\mu\nu}^{(n_\star)}(r,\pi-\theta)
=
\Pi_\mu{}^\alpha
\Pi_\nu{}^\beta
q_{\alpha\beta}^{(n_\star)}(r,\theta),
\label{eq:completion_reflection_tensor}
\end{equation}
where \(\Pi_\theta{}^\theta=-1\) and \(\Pi_\alpha{}^\alpha=1\) for \(\alpha\neq\theta\). In the reduced ansatz with no single-\(\theta\)-index components, this simply requires the coefficient functions in the correction to be even under \(\theta\mapsto\pi-\theta\). A correction that cancels the field-equation residual but violates Eqs. \eqref{eq:completion_axis_cross_terms}--\eqref{eq:completion_reflection_tensor} is therefore not an admissible residual completion.

Gauge fixing is still not enough. The correction could be made artificially large by allowing every stationary and axisymmetric tensor component to vary. To prevent residual completion from becoming an unconstrained fitting procedure, we impose a minimality principle. Let \(\mathscr{A}\) be the full stationary, axisymmetric, reflection-symmetric correction space compatible with the static limit. Let \(\mathscr{A}_{\rm min}\subset\mathscr{A}\) be the smallest subspace that contains the tensorial and multipolar channels present in the leading obstruction. A minimal completion is a solution of Eq. \eqref{5.31} satisfying
\begin{equation}
\left(
q^{(n_\star)},\psi^{(n_\star)}
\right)
\in
\mathscr{A}_{\rm min}.
\label{5.35}
\end{equation}

For example, if the leading obstruction is purely even and quadrupolar at order \(a^2\), then we restrict the metric correction to
\begin{equation}
q_{\mu\nu}^{(2)}(r,\theta)
=
q_{\mu\nu}^{(2,0)}(r)P_0(\cos\theta)
+
q_{\mu\nu}^{(2,2)}(r)P_2(\cos\theta),
\label{5.36}
\end{equation}
and set all axial first-order corrections to zero unless required by the obstruction. If the leading obstruction is axial and dipolar at order \(a\), we instead restrict to the frame-dragging sector,
\begin{equation}
q_{t\phi}^{(1)}(r,\theta)
=
-\omega_1(r)R^2(r)\sin^2\theta,
\label{5.37}
\end{equation}
and $q_{\mu\nu}^{(1)}=0$ for all other components in the chosen gauge.
These restrictions are not universal assumptions. They are examples of how the minimality principle is implemented once the leading obstruction channel has been identified.

It is also useful to distinguish physical homogeneous modes from residual repair modes. Here a homogeneous mode means a solution of the associated linearized completion equation with zero source. Such a mode does not cancel the Newman--Janis obstruction by itself; instead, it represents residual gauge freedom or a physical zero mode such as a shift of mass, charge, or angular momentum. A homogeneous solution satisfying
\begin{equation}
\mathbb{L}_{\rm NJA}^{(n_\star)}
\left[
q_{\rm hom}^{(n_\star)},\psi_{\rm hom}^{(n_\star)}
\right]
=
0
\label{5.38}
\end{equation}
may correspond to a redefinition of mass, charge, angular momentum, or other conserved quantities. Such modes should be fixed by holding the seed parameters fixed, unless the physical problem explicitly allows a charge renormalization. We impose this by requiring
\begin{equation}
\delta M
=
0,
\quad
\delta Q
=
0,
\quad
\delta J
=
0,
\label{5.39}
\end{equation}
for homogeneous modes not forced by the obstruction. The charges in Eq. \eqref{5.39} may be defined by the appropriate asymptotic or covariant phase-space construction for the theory under consideration \cite{Iyer:1994ys}.

We can now state the residual-completion criterion at leading order. A Newman--Janis deformation is residually completable at order \(a^{n_\star}\) within the ansatz class \(\mathscr{A}_{\rm min}\), gauge \(\mathcal{G}_\mu=0\), and chosen boundary conditions if there exists a pair
\begin{equation}
\left(
q^{(n_\star)},\psi^{(n_\star)}
\right)
\in
\mathscr{A}_{\rm min}
\label{5.40}
\end{equation}
such that
\begin{equation}
\boxed{
\begin{array}{rcl}
\mathbb{L}_{\rm NJA}^{(n_\star)}
\left[
q^{(n_\star)},\psi^{(n_\star)}
\right]
&=&
-\mathcal{O}_{\mu\nu}^{(n_\star)},
\\[4pt]
\mathcal{G}_\mu
\left[
q^{(n_\star)}
\right]
&=&
0,
\\[4pt]
\mathfrak{B}
\left[
q^{(n_\star)},\psi^{(n_\star)}
\right]
&=&
0 .
\end{array}
}
\label{5.41}
\end{equation}
Here \(\mathfrak{B}=0\) denotes the selected boundary, axis-regularity, elementary-flatness, equatorial-reflection, and charge-fixing conditions. Thus the completion is not allowed to repair the field equations by introducing an axial defect, a conical singularity, or an equatorial-parity violation.

\section{Source Preservation and Matter-Sector Obstructions} \label{sec6}
\subsection{Stress-tensor eigenvalue structure}
The obstruction tensor introduced in Sec. \ref{sec3} determines whether a Newman--Janis-type output satisfies the metric equations of the target theory. This is not yet enough to determine whether the rotating geometry represents the same physical source as the static seed. In Einstein gravity, any sufficiently regular metric can be assigned an effective stress tensor by writing \(T_{\mu\nu}^{\rm eff}=G_{\mu\nu}/8\pi\). Such a construction is algebraically valid but physically weak. It does not imply that the source is an anisotropic fluid, a nonlinear electromagnetic field, a scalar field, or any other matter model inherited from the static seed. We therefore require a separate source-preservation test.

The term Segre type refers to the algebraic classification of the mixed tensor \(T^\mu{}_\nu\) according to its eigenvalue degeneracies and Jordan structure. In particular, the classification records whether the stress tensor is diagonalizable and whether its eigenvectors have the causal character required for a physical type-I matter source. 
For a given corrected or uncorrected rotating configuration \((g_{\mu\nu},\Psi)\), the stress tensor defines a mixed tensor
\begin{equation}
T^\mu{}_\nu
=
g^{\mu\alpha}T_{\alpha\nu}.
\label{6.1}
\end{equation}
Its algebraic structure is determined by the eigenvalue problem
\begin{equation}
T^\mu{}_\nu v^\nu
=
\lambda v^\mu .
\label{6.2}
\end{equation}
Equations \eqref{6.1} and \eqref{6.2} are the standard starting point for the Hawking--Ellis or Segre--Plebański classification of stress-energy tensors \cite{Hawking:1973uf,Martin-Moruno:2021niw}. The characteristic polynomial is
\begin{equation}
\det
\left(
T^\mu{}_\nu-\lambda\delta^\mu{}_\nu
\right)
=
0 .
\label{6.3}
\end{equation}
The roots \(\lambda_A\) and the corresponding eigenvectors determine the local algebraic type of the source. A physically ordinary type-I stress tensor has one timelike eigenvector and three spacelike eigenvectors. In an orthonormal frame adapted to these eigenvectors, it can be written as
\begin{equation}
T_{\hat a\hat b}
=
{\rm diag}
\left(
\rho,
p_1,
p_2,
p_3
\right),
\label{6.4}
\end{equation}
where \(\rho\) is the energy density and \(p_i\) are the principal pressures. This is the standard type-I canonical form \cite{Hawking:1973uf,Martin-Moruno:2021niw}.

For the static spherical seed of Sec. \(\ref{sec2}\), the stress tensor has the restricted form
\begin{equation}
T_{\hat a\hat b}^{(0)}
=
{\rm diag}
\left(
\rho_0,
p_{r0},
p_{\perp0},
p_{\perp0}
\right). \label{6.5}
\end{equation}
The equality of the two angular pressures follows from spherical symmetry. A rotating deformation may break this degeneracy. If the mixed tensor \(T^\mu{}_\nu(a)\) remains diagonalizable over the real numbers with one timelike eigenvector and three spacelike eigenvectors, then it can be written in a type-I orthonormal frame as
\begin{equation}
T_{\hat a\hat b}(a)
=
{\rm diag}
\left(
\rho(a),
p_1(a),
p_2(a),
p_3(a)
\right). \label{6.6}
\end{equation}
However, this form is not guaranteed. Some matrices cannot be diagonalized; the stress tensor may instead develop a nontrivial Jordan block or complex eigenvalues. Such cases are not ordinary type-I matter sources and are counted here as algebraic source-preservation obstructions. This is why the Segre classification is used: it distinguishes diagonalizable type-I behavior from non-diagonalizable or nonphysical algebraic structures. Such behavior has been emphasized in analyses of the physical interpretation of Newman--Janis rotating systems, where comparing the eigenvalues, eigenvectors, Segre types, and equations of state of the rotating and nonrotating stress tensors is essential \cite{Beltracchi:2021ris,Beltracchi:2021tcx}.

The eigenvalue problem should be evaluated in a specified observer frame. Let \(e_{\hat a}{}^\mu\) be an orthonormal tetrad satisfying
\begin{equation}
g_{\mu\nu}e_{\hat a}{}^\mu e_{\hat b}{}^\nu
=
\eta_{\hat a\hat b},
\quad
\eta_{\hat a\hat b}
=
{\rm diag}(-1,1,1,1).
\end{equation}
The tetrad may be chosen to be locally nonrotating, adapted to the principal stresses, or adapted to the matter model. We therefore avoid referring to a unique \textit{natural} frame. Source preservation is formulated through the eigenvalues and eigenvectors of \(T^\mu{}_\nu\), while tetrads are used only to display the physical interpretation.

The lowest-order change of the mixed stress tensor is
\begin{equation}
T^\mu{}_\nu(a)
=
T^\mu{}_\nu{}^{(0)}
+
a\,\tau^\mu{}_\nu{}^{(1)}
+
a^2\tau^\mu{}_\nu{}^{(2)}
+
O(a^3).
\label{6.10}
\end{equation}
If the unperturbed eigenvalue \(\lambda_A^{(0)}\) is nondegenerate, its first correction is
\begin{equation}
\lambda_A^{(1)}
=
w^{(A)}{}_\mu
\tau^\mu{}_\nu{}^{(1)}
v_{(A)}^\nu ,
\label{6.11}
\end{equation}
where \(v_{(A)}^\mu\) and \(w^{(A)}{}_\mu\) are the right and left eigenvectors normalized by \(w^{(A)}{}_\mu v_{(A)}^\mu=1\). Equation \eqref{6.11} is the standard first-order eigenvalue perturbation formula for a linear operator with a simple eigenvalue \cite{Kato_1995}. Degenerate sectors, such as the angular-pressure degeneracy \(p_{\perp0}=p_{\perp0}\) in Eq. \eqref{6.5}, must be treated by diagonalizing the perturbation inside the degenerate subspace. This is precisely where rotating deformations can split the angular pressures and change the source class.

A first algebraic source-preservation requirement is therefore
\begin{equation}
{\rm Type}
\left[
T^\mu{}_\nu(a)
\right]
=
{\rm Type}
\left[
T^\mu{}_\nu{}^{(0)}
\right]
\quad
\hbox{}.
\label{6.12}
\end{equation}
to the order considered. For type-I seeds this means that the rotating stress tensor must remain diagonalizable over the real numbers with one timelike eigenvector and three spacelike eigenvectors. If Eq. \eqref{6.12} fails, then the rotating metric may still define a formal effective source, but it does not preserve the algebraic matter class of the seed.

\subsection{Segre-type and equation-of-state preservation}
The algebraic type is necessary but not sufficient. A source may remain type I while failing to preserve the equation of state or field-theoretic structure of the seed matter. We therefore distinguish algebraic source preservation from dynamical source preservation.

Let
\begin{equation}
\mathfrak{s}_0
=
{\rm Segre}
\left[
T^\mu{}_\nu{}^{(0)}
\right]
\label{6.13}
\end{equation}
be the Segre type of the static seed. For a type-I anisotropic fluid with two equal tangential pressures, the schematic Segre structure is
\begin{equation}
\mathfrak{s}_0
=
[1,1(11)],
\label{6.14}
\end{equation}
where the parentheses indicate the degeneracy of the two angular pressure eigenvalues. The precise notation depends on the eigenvalue degeneracies and causal character of the eigenvectors, but the essential point is invariant: the seed source possesses a definite algebraic class.

We define algebraic source preservation by
\begin{equation}
{\rm Segre}
\left[
T^\mu{}_\nu(a)
\right]
\in
\mathscr{S}_{\rm allowed}.
\label{6.15}
\end{equation}
Here \(\mathscr{S}_{\rm allowed}\) is not always a singleton. A rotating anisotropic fluid may legitimately split the two tangential pressures, so one may allow
\begin{equation}
[1,1(11)]
\longrightarrow
[1,111],
\label{6.16}
\end{equation}
provided the resulting stress tensor remains type I and the splitting is compatible with the intended rotating matter model. By contrast, a transition to a stress tensor with complex eigenvalues or a nontrivial null Jordan block is not source-preserving for an ordinary anisotropic-fluid seed.

Next suppose that the static seed satisfies an equation-of-state constraint
\begin{equation}
\mathcal{P}_0
\left(
\rho_0,p_{r0},p_{\perp0};\alpha_A
\right)
=
0,
\label{6.17}
\end{equation}
where \(\alpha_A\) denotes fixed parameters of the matter model. For example, a vacuum-energy core satisfies
\begin{equation}
p_{r0}
=
-\rho_0,
\quad
p_{\perp0}
=
-\rho_0,
\label{6.18}
\end{equation}
while many regular black-hole interiors are represented by anisotropic sources for which the radial and tangential pressures satisfy model-dependent relations. Equation \eqref{6.18} is the standard de Sitter-type equation of state used in regular-core interpretations \cite{Dymnikova:1992ux,Bronnikov:2000vy}.

A rotating configuration is equation-of-state preserving if there exists an allowed rotating relation
\begin{equation}
\mathcal{P}_a
\left(
\rho(a),p_1(a),p_2(a),p_3(a);\alpha_A
\right)
=
0
\label{6.19}
\end{equation}
such that
\begin{equation}
\lim_{a\to0}
\mathcal{P}_a
=
\mathcal{P}_0 .
\label{6.20}
\end{equation}
The function \(\mathcal{P}_a\) may equal \(\mathcal{P}_0\), or it may be a controlled rotational deformation of it. This allowance is important because rotation can change the effective thermodynamic relation seen by a comoving or locally nonrotating observer. For sufficiently large rotation, frame dragging may make the fluid motion relativistic, and the appropriate equation of state may differ from the nonrelativistic or static one. In the present work, however, \(a\) is treated perturbatively, so any change from \(\mathcal{P}_0\) to \(\mathcal{P}_a\) must be specified as part of the rotating matter model and must reduce smoothly to the seed relation as \(a\to0\). What is not allowed is to abandon the seed relation and then declare the Einstein tensor to be a new source without specifying the enlarged matter model.

For nonlinear electrodynamics, source preservation is even more restrictive. A nonlinear electromagnetic field with Lagrangian \(\mathcal{L}_{\rm NED}(F)\), where
\begin{equation}
F
=
\frac{1}{4}F_{\mu\nu}F^{\mu\nu},
\label{6.21}
\end{equation}
has stress tensor
\begin{equation}
T_{\mu\nu}^{\rm NED}
=
\frac{1}{4\pi}
\left(
\mathcal{L}_F F_{\mu\alpha}F_\nu{}^\alpha
-
g_{\mu\nu}\mathcal{L}_{\rm NED}
\right),
\quad
\mathcal{L}_F
=
\frac{d\mathcal{L}_{\rm NED}}{dF}.
\label{6.22}
\end{equation}
This is the standard stress tensor for nonlinear electrodynamics in the convention of Eq. \eqref{6.21}, up to overall sign conventions for the action \cite{Sorokin:2021tge,Ayon-Beato:1998hmi,Bronnikov:2000vy}. A Newman--Janis-generated rotating metric is source-preserving in a nonlinear-electrodynamic model only if there exists a real electromagnetic two-form \(F_{\mu\nu}(a)\) and a Lagrangian \(\mathcal{L}_{\rm NED}\) belonging to the intended model class such that Eq. \eqref{6.22} reproduces the stress tensor required by the rotating geometry and the nonlinear Maxwell equations are satisfied.

For scalar-field sources, the same principle applies. For a minimally coupled scalar field with potential \(V(\Phi)\), the standard stress tensor is
\begin{equation}
T_{\mu\nu}^{\Phi}
=
\nabla_\mu\Phi\nabla_\nu\Phi
-
\frac{1}{2}g_{\mu\nu}
\nabla_\alpha\Phi\nabla^\alpha\Phi
-
g_{\mu\nu}V(\Phi).
\label{6.23}
\end{equation}
This form follows from variation of the minimally coupled scalar action with respect to the metric \cite{Wald:1984rg,Carroll_2019}. A rotating metric whose effective stress tensor cannot be written in the form \eqref{6.23}, for any admissible scalar profile and potential, is not a source-preserving scalar-field deformation.

We therefore define source preservation as the simultaneous satisfaction of three conditions:
\begin{equation}
\boxed{
\begin{array}{rcl}
{\rm Segre}
\left[
T^\mu{}_\nu(a)
\right]
&\in&
\mathscr{S}_{\rm allowed},
\\[4pt]
\mathcal{P}_a
\left(
\rho(a),p_i(a);\alpha_A
\right)
&=&
0,
\\[4pt]
T_{\mu\nu}(a)
&\in&
\mathscr{T}_{\rm model}.
\end{array}
}
\label{6.24}
\end{equation}
Here \(\mathscr{T}_{\rm model}\) is the set of stress tensors realizable by the intended matter theory, such as nonlinear electrodynamics, a scalar field, an anisotropic fluid with specified equation of state, or another fixed source model. Equation \eqref{6.24} is stronger than the effective-source identity \(T_{\mu\nu}^{\rm eff}=G_{\mu\nu}/8\pi\). It requires that the rotating geometry remain in the same physical matter sector as the static seed, or in an explicitly specified controlled enlargement of it.

\subsection{Matter-sector obstruction}
We now define the matter-sector obstruction. Let \(T^\mu{}_\nu(a)\) be the mixed stress tensor associated with either the uncorrected Newman--Janis output or a candidate residual completion. Let
\begin{equation}
\lambda_A(a),
\quad
A=0,1,2,3,
\label{6.25}
\end{equation}
be its eigenvalues, with the convention that \(\lambda_0=-\rho\) in the type-I rest frame and \(\lambda_i=p_i\) for spacelike principal directions. The algebraic obstruction is
\begin{equation}
\mathcal{M}_{\rm alg}
=
{\rm Segre}
\left[
T^\mu{}_\nu(a)
\right]
-
\mathscr{S}_{\rm allowed}.
\label{6.26}
\end{equation}
Equation \eqref{6.26} is symbolic: it means that the Segre type of \(T^\mu{}_\nu(a)\) is compared against the allowed set \(\mathscr{S}_{\rm allowed}\). The obstruction vanishes if and only if the Segre type belongs to the allowed set.

The equation-of-state obstruction is
\begin{equation}
\mathcal{M}_{\rm eos}
=
\mathcal{P}_a
\left(
\rho(a),p_1(a),p_2(a),p_3(a);\alpha_A
\right).
\label{6.27}
\end{equation}
If several equation-of-state relations are required, then \(\mathcal{M}_{\rm eos}\) is a vector of residual functions. For a de Sitter-type core, for example, one may take
\begin{equation}
\mathcal{M}_{\rm eos}
=
\left(
p_1(a)+\rho(a),
p_2(a)+\rho(a),
p_3(a)+\rho(a)
\right)
\label{6.28}
\end{equation}
when the physical model demands vacuum-energy isotropy. If the allowed rotating model permits anisotropy, Eq. \eqref{6.28} must be replaced by the appropriate anisotropic relations.

The model-realizability obstruction is defined by asking whether the stress tensor can be represented by the intended matter fields. For nonlinear electrodynamics, this obstruction may be written schematically as
\begin{equation}
\mathcal{M}_{\rm NED}
=
\left(
T_{\mu\nu}^{\rm req}
-
T_{\mu\nu}^{\rm NED}[F_{\alpha\beta},\mathcal{L}_{\rm NED}],
\,
\nabla_\mu
\left(
\mathcal{L}_F F^{\mu\nu}
\right),
\,
\nabla_{[\mu}F_{\nu\rho]}
\right),
\label{6.29}
\end{equation}
where \(T_{\mu\nu}^{\rm req}\) is the stress tensor required by the gravitational field equations. The final two entries are the nonlinear electromagnetic field equation and the Bianchi identity for the field strength. For a scalar model, the analogous obstruction is
\begin{equation}
\mathcal{M}_{\Phi}
=
\left(
T_{\mu\nu}^{\rm req}
-
T_{\mu\nu}^{\Phi}[\Phi,V],
\,
\nabla^\mu\nabla_\mu\Phi
-
\frac{dV}{d\Phi}
\right).
\label{6.30}
\end{equation}
Equations \eqref{6.29} and \eqref{6.30} are not new matter equations; they are residuals of the standard nonlinear electromagnetic and scalar-field equations \cite{Sorokin:2021tge,Wald:1984rg,Carroll_2019}.

The full matter-sector obstruction functional is
\begin{equation}
\boxed{
\mathcal{M}
=
\left(
\mathcal{M}_{\rm alg},
\mathcal{M}_{\rm eos},
\mathcal{M}_{\rm model}
\right).
}
\label{6.31}
\end{equation}
A rotating deformation is source-preserving if
\begin{equation}
\boxed{
\mathcal{M}
=
0 .
}
\label{6.32}
\end{equation}
If \(\mathcal{M}\neq0\), then at least one of the following is true: the stress tensor has the wrong algebraic type, the eigenvalues fail the required equation of state, or the stress tensor cannot be realized by the intended matter fields. In that case, the geometry may still be an exact solution with an effective source, but it is not a source-preserving rotating counterpart of the original seed.

The matter-sector obstruction can also be expanded in the rotation parameter,
\begin{equation}
\mathcal{M}(a)
=
a^{m_\star}\mathcal{M}^{(m_\star)}
+
O(a^{m_\star+1}),
\label{6.33}
\end{equation}
with
\begin{equation}
m_\star
=
\min
\left\{
m:
\mathcal{M}^{(m)}
\neq0
\right\}.
\label{6.34}
\end{equation}
The integer \(m_\star\) is the leading source-obstruction order. It need not coincide with the leading dynamical obstruction order \(n_\star\). A Newman--Janis metric can solve the gravitational field equations at order \(a^n\) while changing the source interpretation at the same or lower order. Conversely, a metric may fail the field equations while preserving the algebraic source class. The obstruction-completion framework therefore keeps the dynamical and matter-sector residuals separate.

A source-preserving residual completion must satisfy both
\begin{equation}
\mathbb{L}_{\rm NJA}^{(n_\star)}
\left[
q^{(n_\star)},\psi^{(n_\star)}
\right]
=
-\mathcal{O}_{\mu\nu}^{(n_\star)}
\label{6.35}
\end{equation}
and
\begin{equation}
\mathcal{M}^{(m_\star)}
\left[
q,\psi
\right]
=
0 .
\label{6.36}
\end{equation}
Equation \eqref{6.35} repairs the field-equation residual. Equation \eqref{6.36} preserves the physical source class. The two conditions are independent. A correction that cancels \(\mathcal{O}_{\mu\nu}^{(n_\star)}\) may still be unacceptable if it forces \(\mathcal{M}^{(m_\star)}\neq0\).

This motivates the following classification at the level of source interpretation. A Newman--Janis deformation is called strongly source-preserving if
\begin{equation}
\mathcal{O}_{\mu\nu}^{\rm dyn}
=
0,
\quad
\mathcal{O}_{I}^{\rm matter}
=
0,
\quad
\mathcal{M}
=
0 .
\label{6.37}
\end{equation}
It is weakly effective-source admissible if
\begin{equation}
\mathcal{O}_{\mu\nu}^{\rm dyn}
=
0
\label{6.38}
\end{equation}
after defining \(T_{\mu\nu}^{\rm eff}=G_{\mu\nu}/8\pi\), but
\begin{equation}
\mathcal{M}
\neq
0 .
\label{6.39}
\end{equation}
Only the first case supports the claim that the rotating geometry is the rotating counterpart of the original matter-supported seed. The second case supports only the weaker statement that the metric is compatible with some effective stress tensor. This distinction is central for regular black holes, nonlinear electrodynamics, scalar-field seeds, and modified-gravity systems generated by Newman--Janis-type prescriptions.

\section{Obstruction Theorems and No-Go Criteria} \label{sec7}
\subsection{Dynamical admissibility theorem}
We now formulate the obstruction-completion framework as a set of precise solvability statements. Let
\begin{equation}
\mathcal{O}_{\mu\nu}^{\rm NJA}(a)
=
a^{n_\star}\mathcal{O}_{\mu\nu}^{(n_\star)}
+
O(a^{n_\star+1})
\label{7.1}
\end{equation}
be the leading Newman--Janis obstruction defined in Sec. \ref{sec3}. Let \(\mathscr{A}\) denote the chosen space of stationary, axisymmetric, reflection-symmetric residual corrections, including both metric and matter-field perturbations. We write an element of this space as
\begin{equation}
X^{(n_\star)}
=
\left(
q_{\mu\nu}^{(n_\star)},
\psi^{(n_\star)}
\right)
\in
\mathscr{A}.
\label{7.2}
\end{equation}
The leading-order residual-completion equation derived in Sec. \ref{sec5} is
\begin{equation}
\mathbb{L}_{\rm NJA}^{(n_\star)}
X^{(n_\star)}
=
-\mathcal{O}^{(n_\star)} ,
\label{7.3}
\end{equation}
where \(\mathbb{L}_{\rm NJA}^{(n_\star)}\) is the linearized completion operator obtained from the metric and matter equations. Equation \eqref{7.3} is an inhomogeneous linear problem. The sign follows from the convention that \(\mathcal{O}_{\mu\nu}^{\rm NJA}\) is the field-equation residual before correction.

We impose a gauge condition and boundary condition,
\begin{equation}
\mathcal{G}_{\mu}
\left[
q^{(n_\star)}
\right]
=
0,
\quad
\mathfrak{B}
\left[
X^{(n_\star)}
\right]
=
0.
\label{7.4}
\end{equation}
The pair \((\mathcal{G}_\mu,\mathfrak{B})\) is part of the definition of the completion problem. Without it, Eq. \eqref{7.3} is underdetermined by infinitesimal diffeomorphisms, homogeneous perturbations, and possible shifts of global charges.

We now state the basic dynamical admissibility result.

\begin{equation}
\boxed{
\begin{minipage}{0.92\linewidth}
\textbf{Theorem 1.}
Let the off-shell Newman--Janis deformation have leading obstruction \(\mathcal{O}^{(n_\star)}\). Fix an ansatz space \(\mathscr{A}\), a gauge condition \(\mathcal{G}_\mu=0\), and boundary conditions \(\mathfrak{B}=0\). A leading-order residual completion exists if and only if
\begin{equation}
-\mathcal{O}^{(n_\star)}
\in
{\rm Im}
\left(
\mathbb{L}_{\rm NJA}^{(n_\star)}
\big|_{\mathscr{A},\mathcal{G},\mathfrak{B}}
\right).
\label{7.5}
\end{equation}
Equivalently, there exists \(X^{(n_\star)}\in\mathscr{A}\) satisfying Eqs. \eqref{7.3} and \eqref{7.4}.
\end{minipage}
}
\end{equation}

The proof is immediate from the definition of the image of a linear operator. If a correction \(X^{(n_\star)}\) solves Eq. \eqref{7.3}, then \(-\mathcal{O}^{(n_\star)}\) is in the image of the gauge- and boundary-restricted operator. Conversely, if \(-\mathcal{O}^{(n_\star)}\) lies in that image, then by definition there exists a correction \(X^{(n_\star)}\) satisfying the leading-order residual-completion equation.

The theorem becomes computationally useful when the image condition is converted into an adjoint-kernel condition. Let \(\mathbb{L}_{\rm NJA}^{(n_\star)\dagger}\) be the formal adjoint with respect to the inner product
\begin{equation}
\langle Y,Z\rangle
=
\int_{\mathcal{D}}
Y^{\mu\nu}Z_{\mu\nu}\,
d\mu_{(0)}
+
\int_{\mathcal{D}}
Y^I Z_I\,
d\mu_{(0)}.
\label{7.7}
\end{equation}
Here \(\mathcal{D}\) is the chosen meridional domain after quotienting by stationarity and axisymmetry, and \(d\mu_{(0)}\) is the measure induced by the seed geometry on that reduced domain. The first term pairs the metric-equation residuals with metric test tensors, while the second term pairs matter-equation residuals with matter-sector test fields. In a concrete gauge and harmonic decomposition, Eq. \eqref{7.7} reduces to the usual radial inner product for the corresponding ordinary differential system, with the appropriate weight factors inherited from \(d\mu_{(0)}\). The boundary conditions \(\mathfrak{B}=0\) are chosen so that integrations by parts do not leave uncancelled boundary terms. Thus Eq. \eqref{7.7} is a formal pairing used to define the adjoint operator and the corresponding Fredholm compatibility condition; it is not intended as a unique physical norm. For a Fredholm operator, the inhomogeneous equation
\begin{equation}
\mathbb{L}X
=
Y
\label{7.8}
\end{equation}
is solvable only if \(Y\) is orthogonal to the kernel of the adjoint. This is the standard Fredholm alternative for linear operators \cite{Kato_1995,1972}. Thus Eq. \eqref{7.3} requires
\begin{equation}
\boxed{
\left\langle
Z,
\mathcal{O}^{(n_\star)}
\right\rangle
=
0
\quad
\hbox{for all}
\quad
Z\in
{\rm Ker}
\left[
\mathbb{L}_{\rm NJA}^{(n_\star)\dagger}
\right].
}
\label{7.9}
\end{equation}
Equation \eqref{7.9} is the adjoint compatibility condition for residual completion.

The adjoint-kernel compatibility condition in Eq. \eqref{7.9} is not special to Einstein gravity. It is the Fredholm compatibility condition for the gauge-fixed linearized completion operator in any theory for which the corresponding boundary-value problem has the required Fredholm structure. In a diffeomorphism-invariant relativistic theory, this algebraic compatibility condition must also be supplemented by the differential Noether identity relating the metric equations and matter equations. In Einstein gravity, the familiar special case is the contracted Bianchi identity,
\begin{equation}
\nabla^\mu G_{\mu\nu}=0 .
\label{7.10}
\end{equation}
In a general diffeomorphism-invariant metric theory, the analogous identity states that the divergence of the metric field equation is related to the matter Euler--Lagrange equations. At the leading obstruction order, this gives the necessary compatibility condition
\begin{equation}
\nabla_{(0)}^\mu
\mathcal{O}_{\mu\nu}^{(n_\star)}
=
0
\label{7.11}
\end{equation}
whenever the matter equations have also been completed at the same order. If this condition fails, then the obstruction cannot be cancelled by a metric correction alone; one must include the appropriate matter correction, enlarge the ansatz, or reject the Newman--Janis deformation within the chosen theory.

Theorem 1 therefore has two complementary interpretations. Algebraically, it says that residual completion exists precisely when the obstruction lies in the range of the completion operator. Analytically, it says that the obstruction must satisfy all compatibility conditions associated with gauge, boundary conditions, adjoint zero modes, and differential identities. This converts the Newman--Janis failure problem into a well-posed solvability problem.

\subsection{No-go criterion for restricted ansatz classes}
Theorem 1 also provides a precise no-go statement. The no-go result is not a claim that no rotating solution exists in the full theory. It is a claim that no correction exists within a specified Newman--Janis-compatible completion space.

Let
\begin{equation}
\mathscr{A}_{\rm min}
\subset
\mathscr{A}
\label{7.12}
\end{equation}
be the minimal correction space selected by the leading obstruction channel. For example, \(\mathscr{A}_{\rm min}\) may contain only the dipolar axial frame-dragging correction at order \(a\), or only the monopolar and quadrupolar even-parity corrections at order \(a^2\). Let
\begin{equation}
\mathbb{L}_{\rm min}^{(n_\star)}
=
\mathbb{L}_{\rm NJA}^{(n_\star)}
\big|_{\mathscr{A}_{\rm min},\mathcal{G},\mathfrak{B}}
\label{7.13}
\end{equation}
be the completion operator restricted to this minimal space.

The restricted no-go criterion is
\begin{equation}
-\mathcal{O}^{(n_\star)}
\notin
{\rm Im}
\left(
\mathbb{L}_{\rm min}^{(n_\star)}
\right)
\label{7.14}
\end{equation}
no leading-order residual completion exists in $\mathscr{A}_{\rm min}$. Equivalently, if there exists an adjoint zero mode \(Z\) such that
\begin{equation}
Z
\in
{\rm Ker}
\left[
\mathbb{L}_{\rm min}^{(n_\star)\dagger}
\right],
\quad
\left\langle
Z,
\mathcal{O}^{(n_\star)}
\right\rangle
\neq
0,
\label{7.15}
\end{equation}
then the leading obstruction cannot be cancelled within the restricted ansatz class. Equation \eqref{7.15} is the practical Fredholm-type obstruction test.

This statement is deliberately scoped. It does not exclude a completion in a larger space
\begin{equation}
\mathscr{A}_{\rm min}
\subsetneq
\mathscr{A}_{\rm ext},
\label{7.16}
\end{equation}
nor does it exclude a rotating solution not connected to the Newman--Janis ansatz. It says only that the selected Newman--Janis-compatible completion space is too small to absorb the obstruction.

There are several ways in which the no-go condition can arise. First, the obstruction may have support in a tensorial component absent from the correction space. If
\begin{equation}
\Pi_{\perp}
\mathcal{O}^{(n_\star)}
\neq
0 ,
\label{7.17}
\end{equation}
where \(\Pi_{\perp}\) projects onto components orthogonal to the image of \(\mathbb{L}_{\rm min}^{(n_\star)}\), then no correction in \(\mathscr{A}_{\rm min}\) can cancel it. Second, the obstruction may have the wrong angular multipole. If the minimal correction allows only \(\ell=0\) and \(\ell=2\) even-parity perturbations but the obstruction contains an uncancelled \(\ell=4\) component, then Eq. \eqref{7.14} holds. Third, the obstruction may violate a conservation or regularity condition, such as Eq. \eqref{7.11}, making it incompatible with the linearized field equations.

A useful way to display the restricted no-go result is to decompose the obstruction into image and cokernel parts,
\begin{equation}
\mathcal{O}^{(n_\star)}
=
\mathcal{O}^{(n_\star)}_{\rm im}
+
\mathcal{O}^{(n_\star)}_{\rm coker},
\label{7.18}
\end{equation}
with
\begin{equation}
\mathcal{O}^{(n_\star)}_{\rm im}
\in
{\rm Im}
\left(
\mathbb{L}_{\rm min}^{(n_\star)}
\right),
\quad
\mathcal{O}^{(n_\star)}_{\rm coker}
\perp
{\rm Im}
\left(
\mathbb{L}_{\rm min}^{(n_\star)}
\right).
\label{7.19}
\end{equation}
Residual completion exists in \(\mathscr{A}_{\rm min}\) only if
\begin{equation}
\boxed{
\mathcal{O}^{(n_\star)}_{\rm coker}
=
0 .
}
\label{7.20}
\end{equation}
If \(\mathcal{O}^{(n_\star)}_{\rm coker}\neq0\), the uncancelled cokernel component is the irreducible Newman--Janis obstruction within the chosen ansatz class.

The no-go criterion also clarifies the role of gauge. A failure produced only by a pure-gauge component is not a physical obstruction. Let
\begin{equation}
q_{\mu\nu}^{\rm gauge}
=
\nabla_\mu\zeta_\nu
+
\nabla_\nu\zeta_\mu .
\label{7.21}
\end{equation}
If the apparent obstruction can be removed by a correction of the form \eqref{7.21}, then it belongs to the gauge orbit and should not be counted as a dynamical no-go. The no-go test must therefore be applied after gauge fixing or to gauge-invariant combinations of the obstruction. This is the same logic used in black-hole perturbation theory, where physical perturbations are separated from coordinate artifacts by gauge-invariant master variables or explicit gauge choices \cite{Regge:1957td,Zerilli:1970se,Moncrief:1974am,Martel:2005ir}.

We can now state the restricted no-go theorem.

\begin{equation}
\boxed{
\begin{minipage}{0.92\linewidth}
\textbf{Theorem 2.}
Fix a Newman--Janis-compatible ansatz class \(\mathscr{A}_{\rm min}\), gauge condition \(\mathcal{G}_\mu=0\), and boundary conditions \(\mathfrak{B}=0\). If the leading obstruction has a nonzero projection onto the cokernel of the restricted completion operator, then no leading-order residual completion exists within \(\mathscr{A}_{\rm min}\). The conclusion does not exclude completion in a larger ansatz class or the existence of a rotating solution unrelated to the Newman--Janis deformation.
\end{minipage}
}
\label{7.22}
\end{equation}

The proof follows from the adjoint-kernel condition in Eq. \eqref{7.15}. If an adjoint zero mode has nonzero overlap with the obstruction, then the obstruction cannot be in the range of the restricted operator. Therefore Eq. \eqref{7.3} has no solution in \(\mathscr{A}_{\rm min}\). The final qualification follows because enlarging the domain of the operator can enlarge its image.

\subsection{Source-preserving obstruction theorem}
The previous theorems address dynamical completion. They do not yet ensure that the completed solution preserves the physical matter sector of the seed. We now combine the residual-completion condition with the matter-sector obstruction of Sec. \ref{sec6}.

Let
\begin{equation}
\mathcal{M}(a)
=
a^{m_\star}\mathcal{M}^{(m_\star)}
+
O(a^{m_\star+1})
\label{7.23}
\end{equation}
be the leading matter-sector obstruction, where \(\mathcal{M}\) includes algebraic, equation-of-state, and model-realizability components. A source-preserving residual completion must solve the coupled system
\begin{equation}
\mathbb{L}_{\rm NJA}^{(n_\star)}
X^{(n_\star)}
=
-\mathcal{O}^{(n_\star)}
\label{7.24}
\end{equation}
and
\begin{equation}
\mathcal{M}^{(m_\star)}
\left[
X^{(n_\star)}
\right]
=
0 ,
\label{7.25}
\end{equation}
to the required perturbative order. The first equation repairs the field-equation residual. The second equation enforces preservation of the seed matter interpretation.

The matter-sector obstruction is independent of the metric-equation obstruction. It is possible that
\begin{equation}
\mathcal{O}_{\mu\nu}^{(n_\star)}
=
0,
\quad
\mathcal{M}^{(m_\star)}
\neq
0 .
\label{7.26}
\end{equation}
In that case, the Newman--Janis output may solve the gravitational field equations with some effective stress tensor, but it does not preserve the source class of the seed. Conversely, one may have
\begin{equation}
\mathcal{O}_{\mu\nu}^{(n_\star)}
\neq
0,
\quad
\mathcal{M}^{(m_\star)}
=
0 ,
\label{7.27}
\end{equation}
meaning that the matter interpretation remains algebraically admissible but the field equations are not yet satisfied. Both cases occur naturally in attempts to interpret Newman--Janis-generated regular black holes and nonlinear-electrodynamic rotating systems. In particular, comparisons of stress-tensor eigenvalues, Segre types, and equations of state have shown that rotating Newman--Janis systems need not preserve the matter interpretation of their static seeds \cite{Beltracchi:2021ris,Beltracchi:2021tcx}.

We define the source-preserving solution space by
\begin{equation}
\mathscr{A}_{\rm src}
=
\left\{
X\in\mathscr{A}_{\rm min}
:
\mathcal{M}[X]=0
\right\}.
\label{7.28}
\end{equation}
The completion operator restricted to this source-preserving space is
\begin{equation}
\mathbb{L}_{\rm src}^{(n_\star)}
=
\mathbb{L}_{\rm NJA}^{(n_\star)}
\big|_{\mathscr{A}_{\rm src},\mathcal{G},\mathfrak{B}} .
\label{7.29}
\end{equation}
The source-preserving admissibility condition is then
\begin{equation}
\boxed{
-\mathcal{O}^{(n_\star)}
\in
{\rm Im}
\left(
\mathbb{L}_{\rm src}^{(n_\star)}
\right).
}
\label{7.30}
\end{equation}
This condition is stronger than Eq. \eqref{7.5} because
\begin{equation}
\mathscr{A}_{\rm src}
\subseteq
\mathscr{A}_{\rm min}
\label{7.31}
\end{equation}
and therefore
\begin{equation}
{\rm Im}
\left(
\mathbb{L}_{\rm src}^{(n_\star)}
\right)
\subseteq
{\rm Im}
\left(
\mathbb{L}_{\rm min}^{(n_\star)}
\right).
\label{7.32}
\end{equation}
A deformation can be dynamically completable but not source-preserving.

We now state the source-preserving obstruction theorem.

\begin{equation}
\boxed{
\begin{minipage}{0.92\linewidth}
\textbf{Theorem 3.}
Let a Newman--Janis-type deformation have leading dynamical obstruction \(\mathcal{O}^{(n_\star)}\) and matter-sector obstruction \(\mathcal{M}^{(m_\star)}\). A source-preserving residual completion exists to the relevant order only if there is a correction \(X\in\mathscr{A}_{\rm min}\) such that
\begin{equation}
\mathbb{L}_{\rm NJA}^{(n_\star)}X
=
-\mathcal{O}^{(n_\star)},
\quad
\mathcal{M}[X]
=
0,
\label{7.33}
\end{equation}
together with the imposed gauge and boundary conditions. Equivalently,
\begin{equation}
-\mathcal{O}^{(n_\star)}
\in
{\rm Im}
\left(
\mathbb{L}_{\rm src}^{(n_\star)}
\right).
\label{7.34}
\end{equation}
If Eq. \eqref{7.5} holds but Eq. \eqref{7.34} fails, then the Newman--Janis deformation is dynamically completable but not source-preserving.
\end{minipage}
}
\label{7.35}
\end{equation}

The proof is again an image-space statement. A source-preserving completion is a completion belonging to \(\mathscr{A}_{\rm src}\). Therefore the obstruction must lie in the image of the completion operator restricted to \(\mathscr{A}_{\rm src}\). If the obstruction lies in the larger image of \(\mathbb{L}_{\rm min}^{(n_\star)}\) but not in the smaller image of \(\mathbb{L}_{\rm src}^{(n_\star)}\), then a dynamical correction exists, but every such correction changes the matter sector.

The theorem gives a precise meaning to the distinction between an effective-source rotating metric and a source-preserving rotating metric. An effective-source completion requires only
\begin{equation}
G_{\mu\nu}[g]
=
8\pi T_{\mu\nu}^{\rm eff}
\label{7.36}
\end{equation}
in Einstein gravity. A source-preserving completion requires, in addition, that
\begin{equation}
T_{\mu\nu}^{\rm eff}
\in
\mathscr{T}_{\rm model},
\quad
{\rm Segre}
\left[
T^\mu{}_\nu
\right]
\in
\mathscr{S}_{\rm allowed},
\quad
\mathcal{P}_a(\rho,p_i)=0 .
\label{7.37}
\end{equation}
The first condition in Eq. \eqref{7.37} enforces realizability by the intended matter theory. The second enforces the allowed algebraic type. The third enforces the equation-of-state constraint. These are precisely the conditions encoded in \(\mathcal{M}=0\).

The final obstruction classification can now be summarized as follows:
\begin{equation}
\begin{array}{lll}
{\rm Class\ I}
:
\mathcal{O}^{\rm NJA}=0,\quad \mathcal{M}=0,
\\[3pt]
{\rm Class\ II}
:
\mathcal{O}^{\rm NJA}\neq0,\quad
-\mathcal{O}^{(n_\star)}
\in
{\rm Im}
\left(
\mathbb{L}_{\rm src}^{(n_\star)}
\right),
\\[3pt]
{\rm Class\ III}
:
-\mathcal{O}^{(n_\star)}
\notin
{\rm Im}
\left(
\mathbb{L}_{\rm min}^{(n_\star)}
\right)
\nn
{\rm or}
\quad
\mathcal{M}\neq0
\ \hbox{for all dynamical completions}.
\end{array}
\label{7.38}
\end{equation}
Class I consists of Newman--Janis outputs that are already admissible. Class II consists of outputs that fail off shell but admit a controlled source-preserving residual completion. Class III consists of outputs that are either dynamically obstructed within the chosen ansatz class or dynamically repairable only by changing the matter sector. The classification is perturbative and ansatz-dependent, but it is sharper than the traditional binary distinction between a Newman--Janis metric that \textit{works} and one that \textit{fails}.

\section{Application: A Non-Kerr ONJA Deformation of Schwarzschild} \label{sec8}
In this section we give a concrete application of the obstruction framework to a Schwarzschild seed with constant mass \(M\), but with a complexification different from the one that produces Kerr. The purpose is not to construct a physically preferable rotating black hole. The purpose is to show explicitly how the obstruction tensor detects a failed Newman--Janis output and how the leading obstruction labels \(n_\star\) and \(\ell_\star\) are computed.

We work in vacuum Einstein gravity. Hence the target equations are
\begin{equation}
G_{\mu\nu}=0,
\label{8.1}
\end{equation}
and the Newman--Janis obstruction tensor reduces to
\begin{equation}
\mathcal{O}_{\mu\nu}
=
G_{\mu\nu}
\left[
g^{\rm trial}
\right].
\label{8.2}
\end{equation}
The static seed is Schwarzschild,
\begin{equation}
F(r)=1-\frac{2M}{r},
\quad
R(r)=r,
\label{8.3}
\end{equation}
written in the outgoing Eddington--Finkelstein form
\begin{equation}
ds_0^2
=
-
\left(
1-\frac{2M}{r}
\right)du^2
-
2\,du\,dr
+
r^2 d\Omega_2^2 .
\label{8.4}
\end{equation}

For the usual Kerr-generating ONJA prescription, the Schwarzschild factor is complexified as
\begin{equation}
\frac{1}{r}
\longmapsto
\frac{r}{r\bar r}
=
\frac{r}{r^2+a^2\cos^2\theta},
\label{8.5}
\end{equation}
so that
\begin{equation}
F(r)
\longmapsto
F_{\rm K}(r,\theta;a)
=
1-\frac{2Mr}{\Sigma},
\quad
\Sigma=r^2+a^2\cos^2\theta .
\label{8.6}
\end{equation}
This gives the Kerr metric.

Instead, we deliberately choose the alternative ONJA complexification
\begin{equation}
\frac{1}{r}
\longmapsto
\frac{1}{\sqrt{r\bar r}}
=
\frac{1}{\sqrt{r^2+a^2\cos^2\theta}},
\label{8.7}
\end{equation}
which gives
\begin{equation}
F(r)
\longmapsto
F_{\rm alt}(r,\theta;a)
=
1-
\frac{2M}{\sqrt{\Sigma}},
\quad
\Sigma=r^2+a^2\cos^2\theta .
\label{8.8}
\end{equation}
This is an explicit ONJA-type complexification, but it is not the Kerr complexification. Following the ONJA tetrad procedure, we now insert \(F_{\rm alt}\) into the transformed null tetrad, reconstruct the inverse metric using Eq.~\eqref{eq:onja_inverse_metric_reconstruction}, and then invert the result to obtain the covariant metric. This gives the following real Lorentzian trial metric. The corresponding Newman--Janis trial metric in outgoing Eddington--Finkelstein-type coordinates is
\begin{equation}
\begin{aligned}
&ds_{\rm alt}^2
={}
-
F_{\rm alt}\,du^2
-
2\,du\,dr
+
2a\sin^2\theta
\left(
F_{\rm alt}-1
\right)du\,d\phi
\nn&+
2a\sin^2\theta\,dr\,d\phi
+
\Sigma\,d\theta^2
+
\sin^2\theta
\left[
\Sigma
+
a^2\sin^2\theta
\left(
2-F_{\rm alt}
\right)
\right]d\phi^2 .
\end{aligned}
\label{8.9}
\end{equation}
For comparison, replacing \(F_{\rm alt}\) in Eq. \eqref{8.9} by \(F_{\rm K}\) gives the Kerr metric in the same outgoing coordinate convention.

The two complexifications agree in the static limit but differ at order \(a^2\):
\begin{equation}
F_{\rm K}
=
1-\frac{2M}{r}
+
\frac{2Ma^2\cos^2\theta}{r^3}
+
O(a^4),
\label{8.10}
\end{equation}
whereas
\begin{equation}
F_{\rm alt}
=
1-\frac{2M}{r}
+
\frac{Ma^2\cos^2\theta}{r^3}
+
O(a^4).
\label{8.11}
\end{equation}
Thus the alternative ONJA metric agrees with Kerr through first order in \(a\), but differs from Kerr at order \(a^2\). Since Kerr is vacuum, any nonzero Einstein tensor of Eq. \eqref{8.9} measures directly the obstruction generated by the alternative complexification.

Substituting Eq. \eqref{8.9} into the Einstein tensor and expanding in powers of \(a\), one finds
\begin{equation}
G_{\mu\nu}
\left[
g_{\rm alt}
\right]
=
a^2\mathcal{O}_{\mu\nu}^{(2)}
+
O(a^3),
\label{8.12}
\end{equation}
with no order-\(a\) contribution:
\begin{equation}
\mathcal{O}_{\mu\nu}^{(1)}=0.
\label{8.13}
\end{equation}
The nonzero components of \(\mathcal{O}_{\mu\nu}^{(2)}\) are
\begin{equation}
\mathcal{O}_{uu}^{(2)}
=
\frac{M}{r^6}
\left(
4M\cos^2\theta
-
r\sin^2\theta
\right),
\label{8.14}
\end{equation}
\begin{equation}
\mathcal{O}_{ur}^{(2)}
=
-\frac{2M}{r^5}\cos^2\theta,
\label{8.15}
\end{equation}
\begin{equation}
\mathcal{O}_{u\theta}^{(2)}
=
-\frac{3M}{2r^4}\sin 2\theta,
\label{8.16}
\end{equation}
\begin{equation}
\mathcal{O}_{\theta\theta}^{(2)}
=
-\frac{3M}{r^3}\cos^2\theta,
\label{8.17}
\end{equation}
and
\begin{equation}
\mathcal{O}_{\phi\phi}^{(2)}
=
-\frac{3M}{r^3}
\sin^2\theta\cos^2\theta .
\label{8.18}
\end{equation}
All other components vanish at this order, up to the symmetry
\(\mathcal{O}_{\mu\nu}^{(2)}=\mathcal{O}_{\nu\mu}^{(2)}\).

Equations \eqref{8.12}--\eqref{8.18} show that the first nonzero dynamical obstruction occurs at order \(a^2\). Therefore, according to Eq. \eqref{3.25},
\begin{equation}
n_\star=2.
\label{8.19}
\end{equation}
The angular content can be read off, for example, from Eq. \eqref{8.15}. Since
\begin{equation}
\cos^2\theta
=
\frac{1}{3}
\left[
1+2P_2(\cos\theta)
\right],
\label{8.20}
\end{equation}
we have
\begin{equation}
\mathcal{O}_{ur}^{(2)}
=
-\frac{2M}{3r^5}
\left[
1+2P_2(\cos\theta)
\right].
\label{8.21}
\end{equation}
Thus the leading obstruction contains a nonzero monopolar part. With the definition of Eq. \eqref{3.30}, the first angular channel is therefore
\begin{equation}
\ell_\star=0.
\label{8.22}
\end{equation}
The same obstruction also contains quadrupolar even-parity pieces. Thus the leading dynamical obstruction channel is
\begin{equation}
(n_\star,\ell_\star)=(2,0),
\label{8.23}
\end{equation}
with additional \(\ell=2\) structure at the same order.

We next test the geometrical obstruction channels. The trial metric \eqref{8.9} is regular on the axis in the elementary-flatness sense. Indeed,
\begin{equation}
g_{\phi A}=O(\sin^2\theta),
\quad
A\neq \phi,
\quad
\theta\to0,\pi,
\label{8.24}
\end{equation}
and
\begin{equation}
\lim_{\theta\to0,\pi}
\frac{g_{\phi\phi}}
{g_{\theta\theta}\sin^2\theta}
=
1.
\label{8.25}
\end{equation}
Hence
\begin{equation}
\mathcal{A}_+=0,
\quad
\mathcal{A}_-=0.
\label{8.26}
\end{equation}
The metric is also invariant under equatorial reflection because \(F_{\rm alt}\) and \(\Sigma\) depend on \(\cos^2\theta\). Therefore
\begin{equation}
\mathcal{R}_{\rm eq}=0.
\label{8.27}
\end{equation}

However, the metric fails the circularity/Boyer--Lindquist sector at order \(a^2\). With \(\xi^\mu=\partial_u\) and \(\eta^\mu=\partial_\phi\), one finds, up to the positive volume-form normalization,
\begin{equation}
\mathcal{C}_{\xi}
=
\frac{2Ma^2}{r^3}
\sin^2\theta\cos\theta
+
O(a^4),
\quad
\mathcal{C}_{\eta}
=
O(a^4).
\label{8.28}
\end{equation}
Thus the alternative complexification produces a non-circular obstruction at order \(a^2\).

The same conclusion appears from the restricted Boyer--Lindquist integrability test. Using Eqs. \eqref{4.19} and \eqref{4.20}, the candidate radial shifts obey
\begin{equation}
\mathcal{B}_{\lambda}
\equiv
\partial_\theta\lambda
=
\frac{Ma^2\sin 2\theta}
{r(r-2M)^2}
+
O(a^4),
\label{8.29}
\end{equation}
and
\begin{equation}
\mathcal{B}_{\chi}
\equiv
\partial_\theta\chi
=
O(a^3).
\label{8.30}
\end{equation}
Therefore the alternative trial metric does not admit the simple Boyer--Lindquist radial transformation assumed in Eq. \eqref{4.13} at the first obstructed order. The leading geometrical obstruction order is also
\begin{equation}
n_{\rm geom}=2.
\label{8.31}
\end{equation}

Since the target source is vacuum, source preservation is particularly simple. The seed has
\begin{equation}
T_{\mu\nu}^{(0)}=0.
\label{8.32}
\end{equation}
For the alternative rotating metric, however,
\begin{equation}
G_{\mu\nu}[g_{\rm alt}]
=
a^2\mathcal{O}_{\mu\nu}^{(2)}
+
O(a^3)
\neq0.
\label{8.33}
\end{equation}
Thus the trial metric can be interpreted only as an effective-source geometry unless one modifies the metric. It is not a source-preserving vacuum deformation of Schwarzschild.

Finally, we show how residual completion works in this example. The leading difference between the Kerr-producing complexification and the alternative complexification is
\begin{equation}
F_{\rm K}-F_{\rm alt}
=
\frac{Ma^2\cos^2\theta}{r^3}
+
O(a^4).
\label{8.34}
\end{equation}
At order \(a^2\), this corresponds to the metric correction
\begin{align}
&\delta g_{uu}
=
-\frac{Ma^2\cos^2\theta}{r^3},
\nn
&\delta g_{\mu\nu}=0
\quad
\hbox{for all other components at order }a^2.
\label{8.35}
\end{align}
Equivalently,
\begin{equation}
q_{uu}^{(2)}
=
-\frac{M}{r^3}\cos^2\theta
=
-\frac{M}{3r^3}
\left[
1+2P_2(\cos\theta)
\right].
\label{8.36}
\end{equation}
Adding Eq. \eqref{8.35} converts \(F_{\rm alt}\) into \(F_{\rm K}\) through order \(a^2\). Since the resulting metric is Kerr through this order, the residual-completion equation gives
\begin{equation}
\delta G_{\mu\nu}
\left[
q^{(2)}
\right]
=
-\mathcal{O}_{\mu\nu}^{(2)}.
\label{8.37}
\end{equation}
Therefore the leading obstruction lies in the image of the completion operator once the even-parity monopole-plus-quadrupole correction \eqref{8.36} is allowed.

The example above demonstrates the obstruction-completion procedure explicitly. The wrong ONJA complexification agrees with the Kerr complexification through order \(a\), but fails at order \(a^2\). The failure is detected simultaneously by the vacuum Einstein obstruction, by the circularity/Boyer--Lindquist obstruction, and by the source-preservation test. The obstruction labels are
\begin{equation}
\boxed{
n_\star=2,
\quad
\ell_\star=0,
}
\label{8.38}
\end{equation}
with additional quadrupolar structure at the same order. The minimal correction \eqref{8.36} restores the Kerr complexification at the first obstructed order. Thus the framework does not merely state that a Newman--Janis output may fail; it identifies the order, angular channel, obstruction tensor, geometrical failure, source failure, and the correction needed to remove the leading residual.

\section{Conclusions} \label{sec9}
We have formulated an obstruction and residual-completion framework for Newman--Janis-type deformations. The central idea is to treat the Newman--Janis algorithm, in general, as an off-shell ansatz-generating map rather than as a solution-generating theorem. Its output must therefore be tested against the field equations, matter equations, geometrical admissibility conditions, and the source interpretation inherited from the static seed.

The Newman--Janis obstruction tensor turns the failure of a rotating trial configuration into a computable residual. If it vanishes, the trial geometry is dynamically admissible; if not, its leading order and angular channel identify where a residual correction must begin. Separating this obstruction into dynamical, geometrical, coordinate-integrability, and matter-sector channels also clarifies that a Kerr-like metric form or an effective stress tensor alone does not guarantee circularity, Boyer--Lindquist integrability, matter-field closure, or source preservation.

Residual completion converts this failure into a controlled solvability problem. At the leading obstruction order, the correction must solve a linearized equation sourced by the negative of the obstruction tensor. Solvability requires the obstruction to lie in the image of a gauge-fixed completion operator subject to boundary, regularity, minimality, and source-sector constraints. A nonzero projection onto the corresponding cokernel gives a no-go criterion within the chosen ansatz class.

The Schwarzschild example shows how the framework works in practice. A deliberately non-Kerr ONJA complexification agrees with the Kerr-generating complexification through first order in the rotation parameter, but it develops a nonzero obstruction at order \(a^2\). The explicit computation gives \(n_\star=2\) and \(\ell_\star=0\), with additional quadrupolar structure at the same order. The same example also exhibits a circularity/Boyer--Lindquist obstruction and a failure of vacuum source preservation. The residual completion is the minimal even-parity correction that restores the Kerr complexification at order \(a^2\). This demonstrates that the obstruction formalism is not only a consistency-check framework but also constructive.

The framework therefore replaces the question \textit{Which complexification is correct}? with a sharper test: whether the Newman--Janis obstruction vanishes or can be cancelled by a controlled completion preserving the intended geometry and matter sector. Future extensions include higher-order completion, nonlinear-electrodynamic and scalar-field examples, hidden-symmetry constraints, stability, and observational tests.

\acknowledgments
R. P. and A. \"O.  would like to acknowledge networking support of the COST Action CA21106 - COSMIC WISPers in the Dark Universe: Theory, astrophysics and experiments (CosmicWISPers), the COST Action CA22113 - Fundamental challenges in theoretical physics (THEORY-CHALLENGES), the COST Action CA21136 - Addressing observational tensions in cosmology with systematics and fundamental physics (CosmoVerse), the COST Action CA23130 - Bridging high and low energies in search of quantum gravity (BridgeQG), and the COST Action CA23115 - Relativistic Quantum Information (RQI) funded by COST (European Cooperation in Science and Technology). A. \"O. also thanks to EMU, TUBITAK, ULAKBIM (Turkiye) and SCOAP3 (Switzerland) for their support.

\bibliography{ref}

\end{document}